\documentclass[useAMS,usenatbib]{mnras}

\usepackage{graphicx}	

\usepackage{deluxetable}

\title{Investigating the physical properties of transiting hot Jupiters with the 1.5-m Kuiper Telescope}
\author[Turner et al.]
  {Jake D. Turner$^{1,2}$,
  Robin M. Leiter$^{1}$,
  Lauren I. Biddle$^{2,3}$, 
  Kyle A. Pearson$^{4}$,
   \newauthor 
    Kevin K. Hardegree-Ullman$^{5}$,
    Robert M. Thompson$^{2}$, 
    Johanna K. Teske$^{2,6}$, 
     \newauthor 
    Ian T. Cates$^{2}$,
    Kendall L. Cook$^{2}$, 
    Michael P. Berube$^{2}$, 
    Megan N. Nieberding$^{2}$,
     \newauthor 
    Christen K. Jones$^{2}$, 
    Brandon Raphael$^{2}$, 
    Spencer Wallace$^{2}$, 
     Zachary T. Watson$^{2}$, 
     \newauthor 
    Robert E. Johnson$^{7}$\\
  $^1$Department of Astronomy, University of Virginia, Charlottesville, VA 22904, USA \\
  $^2$Steward Observatory, University of Arizona, Tucson, AZ, 85721, USA\\
  $^3$Northern Arizona University, Department of Physics $\&$ Astronomy, Flagstaff, AZ \\
  $^4$Lunar and Planetary Laboratory, University of Arizona, Tucson, AZ, 85721, USA	\\
  $^5$Department of Physics and Astronomy, University of Toledo, Toledo, OH, 43606, USA \\
  $^6$Currently an Origins Fellow at Carnegie DTM/OCIW\\
  $^7$Department of Materials Science $\&$ Engineering, University of Virginia, Charlottesville, VA 22904, USA\\
 }
\pagerange{\pageref{firstpage}--\pageref{lastpage}} \pubyear{2016}
\begin{document}
\setcounter{equation}{0}
\label{firstpage}
\maketitle
\begin{abstract}
We present new photometric data of 11 hot Jupiter transiting exoplanets (CoRoT-12b, HAT-P-5b, HAT-P-12b, HAT-P-33b, HAT-P-37b, WASP-2b, WASP-24b, WASP-60b, WASP-80b, WASP-103b, XO-3b) in order to update their planetary parameters and to constrain information about their atmospheres. These observations of CoRoT-12b, HAT-P-37b and WASP-60b are the first follow-up data since their discovery. Additionally, the first near-UV transits of WASP-80b and WASP-103b are presented. We compare the results of our analysis with previous work to search for transit timing variations (TTVs) and a wavelength dependence in the transit depth. TTVs may be evidence of a third body in the system and variations in planetary radius with wavelength can help constrain the properties of the exoplanet's atmosphere. For WASP-103b and XO-3b, we find a possible variation in the transit depths that may be evidence of scattering in their atmospheres. The B-band transit depth of HAT-P-37b is found to be smaller than its near-IR transit depth and such a variation may indicate TiO/VO absorption. These variations are detected from 2-4.6$\sigma$, so follow-up observations are needed to confirm these results. Additionally, a flat spectrum across optical wavelengths is found for 5 of the planets (HAT-P-5b, HAT-P-12b, WASP-2b, WASP-24b, WASP-80b), suggestive that clouds may be present in their atmospheres. We calculate a refined orbital period and ephemeris for all the targets, which will help with future observations. No TTVs are seen in our analysis with the exception of WASP-80b and follow-up observations are needed to confirm this possible detection.

\end{abstract}


\section{Introduction} \label{intro}
To date, over 3400 exoplanets have been discovered (NASA Exoplanet Archive; \citealt{Akeson2013}) and most of these planets have been found using the transit method (e.g. \citealt{Charbonneau2000}; \citealt{Henry2000}) in large-scale transit surveys such as \textit{Kepler} (\citealt{Borucki2010}), \textit{K2} (\citealt{Howell2014}), \textit{WASP} (\citealt{Pollacco2006}; \citealt{Cameron2007}) and \textit{CoRoT} (\citealt{Baglin2003}; \citealt{Moutou2013}). Transiting exoplanet systems (TEPs) are of great interest because their radius can be directly measured in relation to their star with photometric observations (\citealt{Charbonneau2000}; \citealt{Henry2000}). With the addition of spectroscopic and radial velocity measurements, many physical properties of TEP systems (mass, radius, semi-major axis, gravity, temperature, eccentricity, orbital period) can be directly measured (e.g., \citealt{Charbonneau2007}). Additionally, multiple-band photometry of a TEP system can be used to constrain the composition of an exoplanet's atmosphere (\citealt{Seager2000}; \citealt{Brown2001}; \citealt{Hubbard2001}; \citealt{Charbonneau2002}). The absorption properties of different species in a planetary atmosphere vary with wavelength, causing an observable variation in the planet's radius. Photometric light curve analysis can also be used to search for transit timing variations (TTVs). TTVs can indicate additional bodies in a TEP system or an unstable orbit caused by tidal forces from the star (e.g., \citealt{Miralda2002}; \citealt{Holman2005}; \citealt{Holman2010}). 

In this work, we present new ground-based photometric data of 11 confirmed transiting hot Jupiter exoplanets. We describe and perform TEP modeling techniques (Section \ref{sec:obs_redu}--\ref{sec:exomop}) to determine the orbital and physical parameters of each system, and compare our results with previous published results to confirm and improve the planetary parameters (Section \ref{sec:physical_properites}--\ref{sec:indiv_systems}). For each system, we combine our results with previous work to search for a variation in planetary radius with wavelength (Section \ref{sec:discuss}), which could indicate Rayleigh scattering, the presence of an absorptive atmosphere, or clouds. Finally, we combine our mid-transit data with previous observations to recalculate each system's orbital period and search for TTVs.

\begin{table*}
\centering
\caption{Journal of observations}
\begin{tabular}{cccccccccc}
\hline 
\hline
Planet 		&Date             & Filter$^{1}$	  & Cadence  	& N$_{pts}$ 		&   OoT RMS$^{2}$ & Res RMS$^{3}$ 	& Seeing  	& k$^{a}$ &  $\chi_{r}^2$ $^{b}$ \\
Name 		& (UT)            &	                          &(s)  &			&(mmag) & (mmag) 	& ($\arcsec$)  \\
\hline
\hline 
CoRoT-12b & 2013 February 15    & R     & 60.39     &265& 6.06 & 4.99 & 1.96-4.06 &4 & 1.32 \\
HAT-P-5b  & 2015 June 6         & U     & 80.64     &191& 3.02 & 2.84 & 1.36-2.17 &4 &0.97 \\
HAT-P-12b & 2014 January 19     & B     & 188.25	&89 & 1.13 & 1.27 & 2.22-2.96 &7 &1.99 \\
HAT-P-33b & 2012 April 6        & R     & 24.29     &706& 3.65 & 4.15 & 1.01-2.68 &4 &7.84 \\
HAT-P-37b & 2015 July 1         & B     & 115.51    &128& 2.37 & 2.88 & 0.98-2.45 &4 &1.10 \\
HAT-P-37b & 2015 July 1         & R     & 115.51    &129& 2.33 & 2.30 & 0.98-2.45 &4 &1.33 \\
WASP-2b   & 2014 June 14        & B     & 59.31     &230& 2.41 & 2.28 & 1.43-2.68 &7 &3.55  \\
WASP-24b  & 2012 March 23       & R     & 24.95     &531& 2.41 & 2.55  & 1.03-2.12 &7 &4.06  \\
WASP-24b  & 2012 April 6        & R     & 26.77     &434&5.14 & 5.73 & 1.01-2.72 &6 &2.04 \\
WASP-60b  & 2012 December 1     & B     & 20.46     &751& 5.30 & 4.65 & 0.83-6.82 &4 &1.21 \\
WASP-80b  & 2014 June 16        & U     & 93.50     &160& 7.76 & 7.20 & 1.54-2.92 &7 &1.15  \\
WASP-103b & 2015 June 3         & U     & 65.05     &363& 3.66 & 3.62 & 1.37-2.56 &6 &1.29 \\
XO-3b     & 2012 November 30    & B     & 40.63     &418& 2.04 & 2.20 & 1.44-2.23 &4 &1.96  \\
\hline
\end{tabular}
\vspace{-2em}
\tablenotetext{1}{Filter: B is the Harris B (330--550 nm), R is the Harris R (550--900 nm) and U is the Bessell U (303--417 nm) } 
\tablenotetext{2}{Out-of-Transit (OoT) root-mean-squared (RMS) relative flux} \\
\tablenotetext{3}{Residual (res) RMS flux after subtracting the EXOplanet Modeling Package (\texttt{EXOMOP}) best-fitting model from the data}
\tablenotetext{a}{$k$ is the degrees of freedom used in the \texttt{EXOMOP} best-fitting model}
\tablenotetext{b}{Reduced $\chi^2$ ($\chi_{r}^2$) calculated using the \texttt{EXOMOP} best-fitting model, $N_{pts}$, and $k$.}
\label{tb:obs_new}
\end{table*}

\section{Observations and Data Reduction} \label{sec:obs_redu}
All the observations were performed at the University of Arizona's Steward Observatory 1.55-m Kuiper Telescope on Mt. Bigelow near Tucson, Arizona. The Mont4k CCD has a field of view of 9.7'$\times$9.7' and contains a 4096$\times$4096 pixel sensor. The CCD is binned 3$\times$3 to achieve a resolution of 0.43$\arcsec$/pixel and binning reduces the read-out time to $\sim$10 s. Our observations were taken with the Bessell U (303--417 nm), Harris B (360--500 nm), and Harris R (550--900 nm) photometric band filters. To ensure accurate timing in these observations, the clocks were synchronized with a GPS every few seconds. In all the data sets, the average shift in the centroid of our targets is less than 0.6 pixels (0.26$\arcsec$) due to excellent autoguiding (the maximum is 3.4 pixels). This telescope has been used extensively in exoplanet transit studies (\citealt{Dittmann2009a,Dittmann2009b,Dittmann2010,Dittmann2012}; \citealt{Scuderi2010}; \citealt{Turner2013a}; \citealt{Teske2013}; \citealt{Pearson2014}; \citealt{Biddle2014}; \citealt{Zellem2015}; \citealt{Turner2016}). A summary of all our observations are displayed in Table \ref{tb:obs_new}.

\begin{figure*}
\centering
\begin{tabular}{cc}
\vspace{0.5cm}
\includegraphics[width=0.33\linewidth,angle=90]{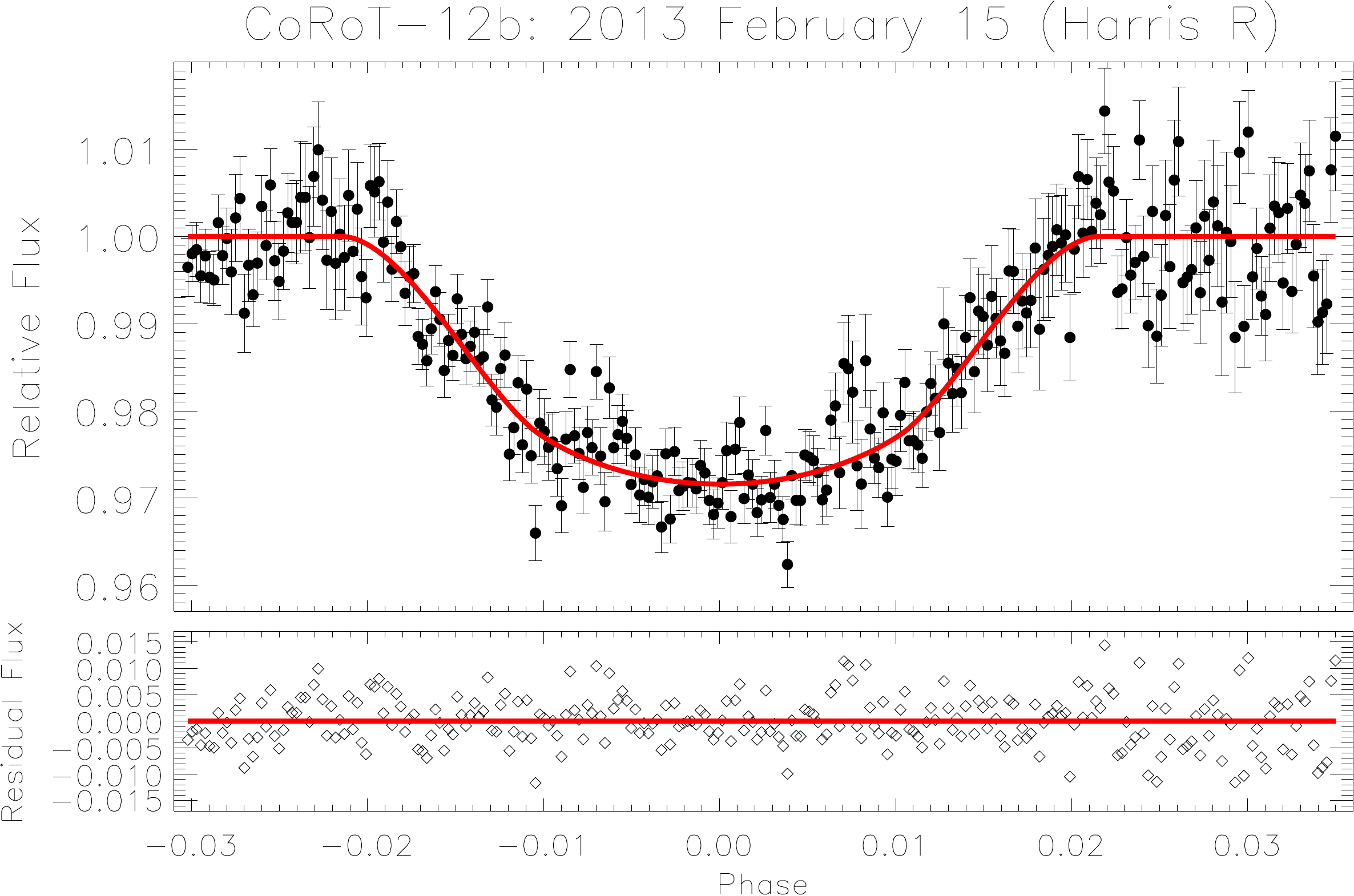} & \includegraphics[width=0.33\linewidth,angle=90]{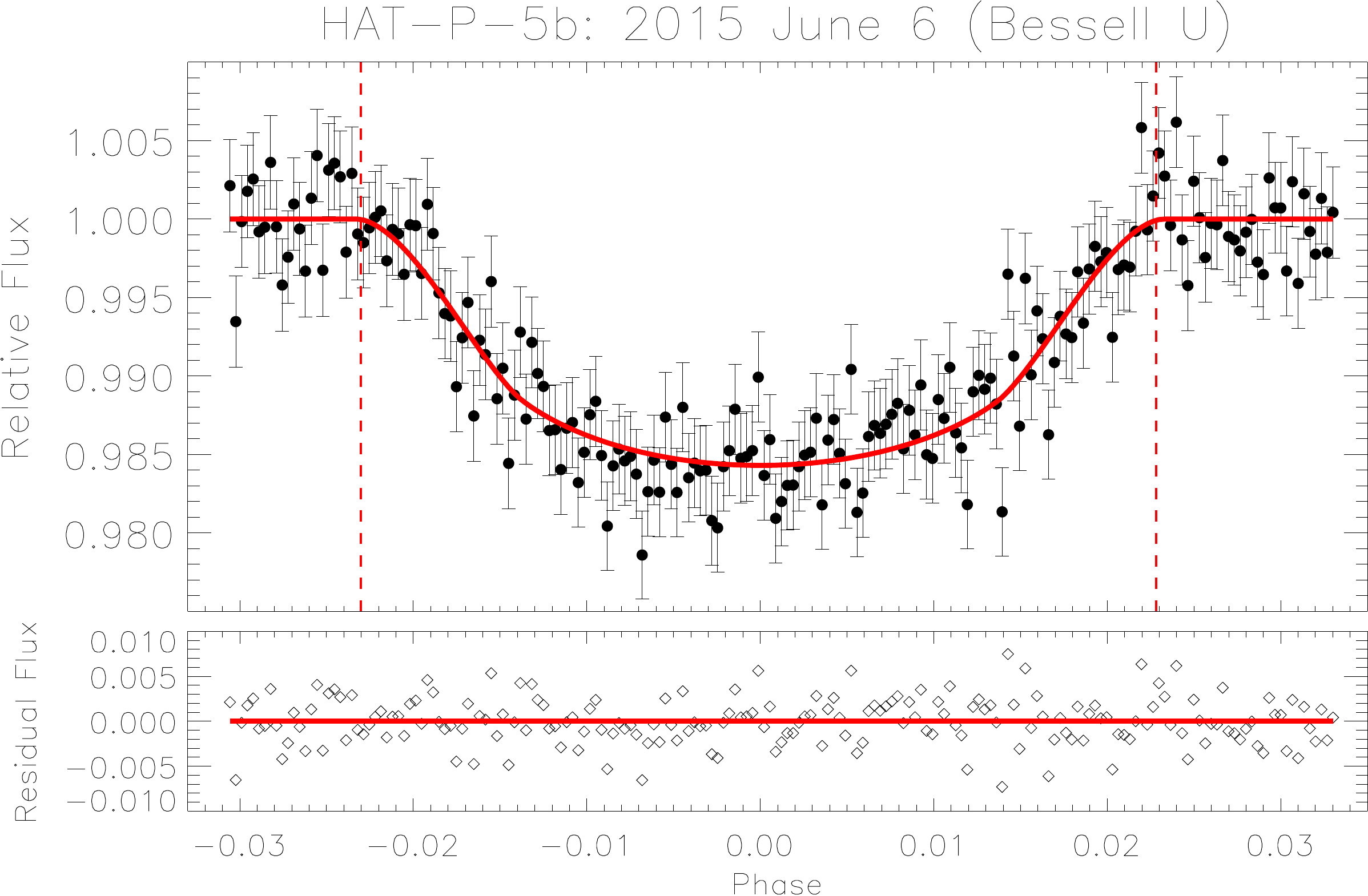}\\
\vspace{0.5cm}
\includegraphics[width=0.33\linewidth,angle=90]{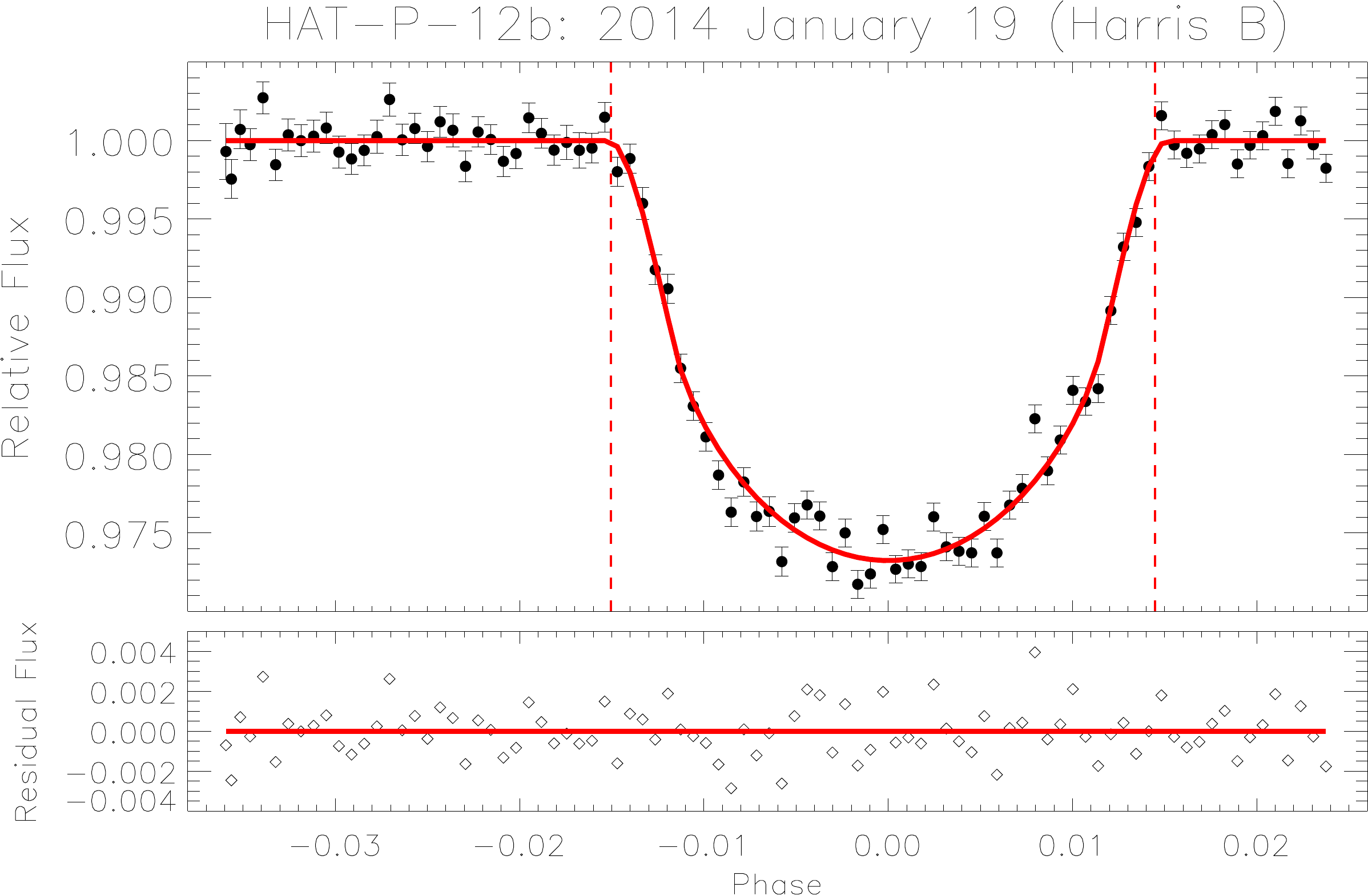} & 
\includegraphics[width=0.33\linewidth,angle=90]{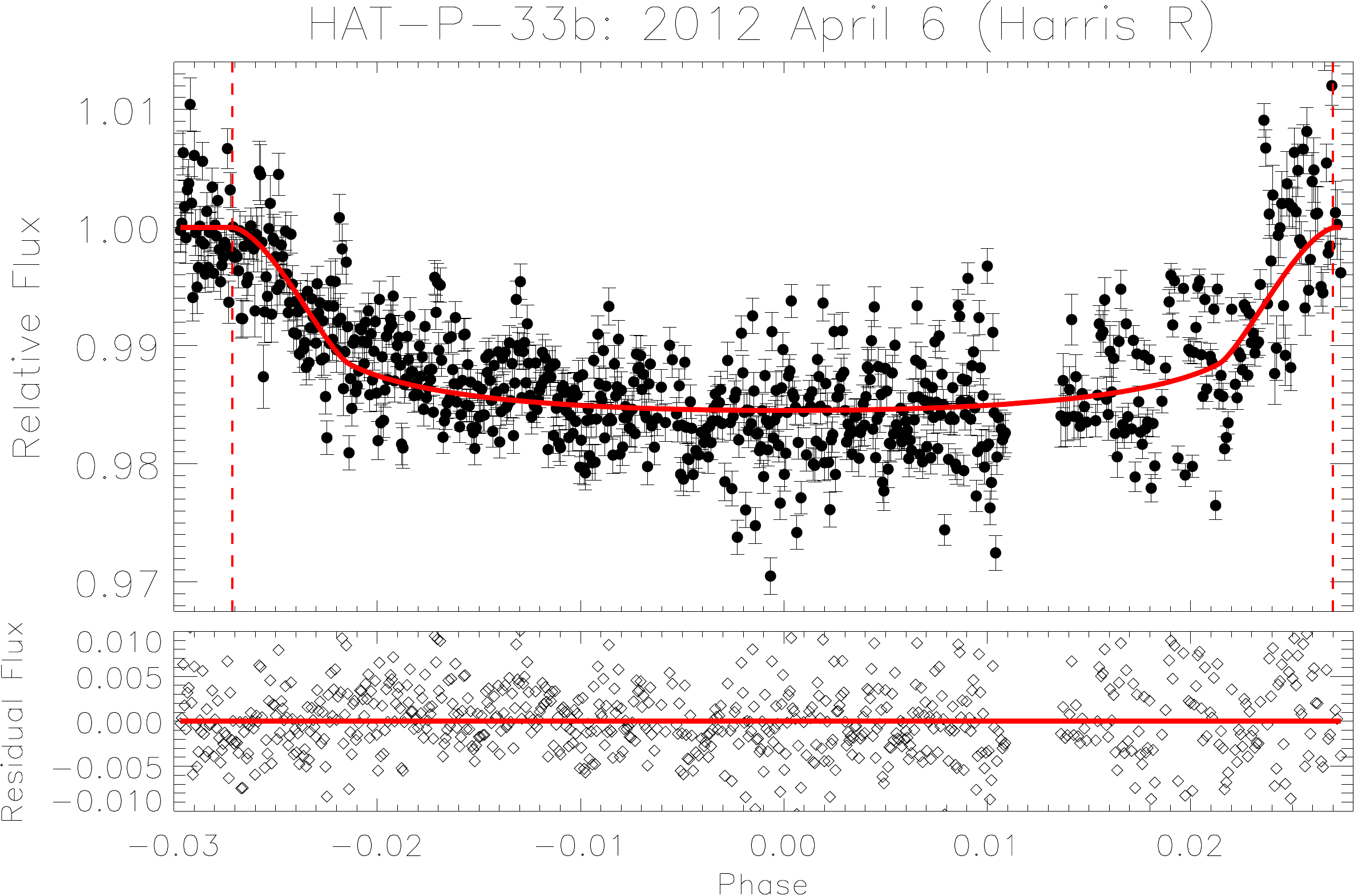} \\ 
\vspace{0.5cm}
\includegraphics[width=0.33\linewidth,angle=90]{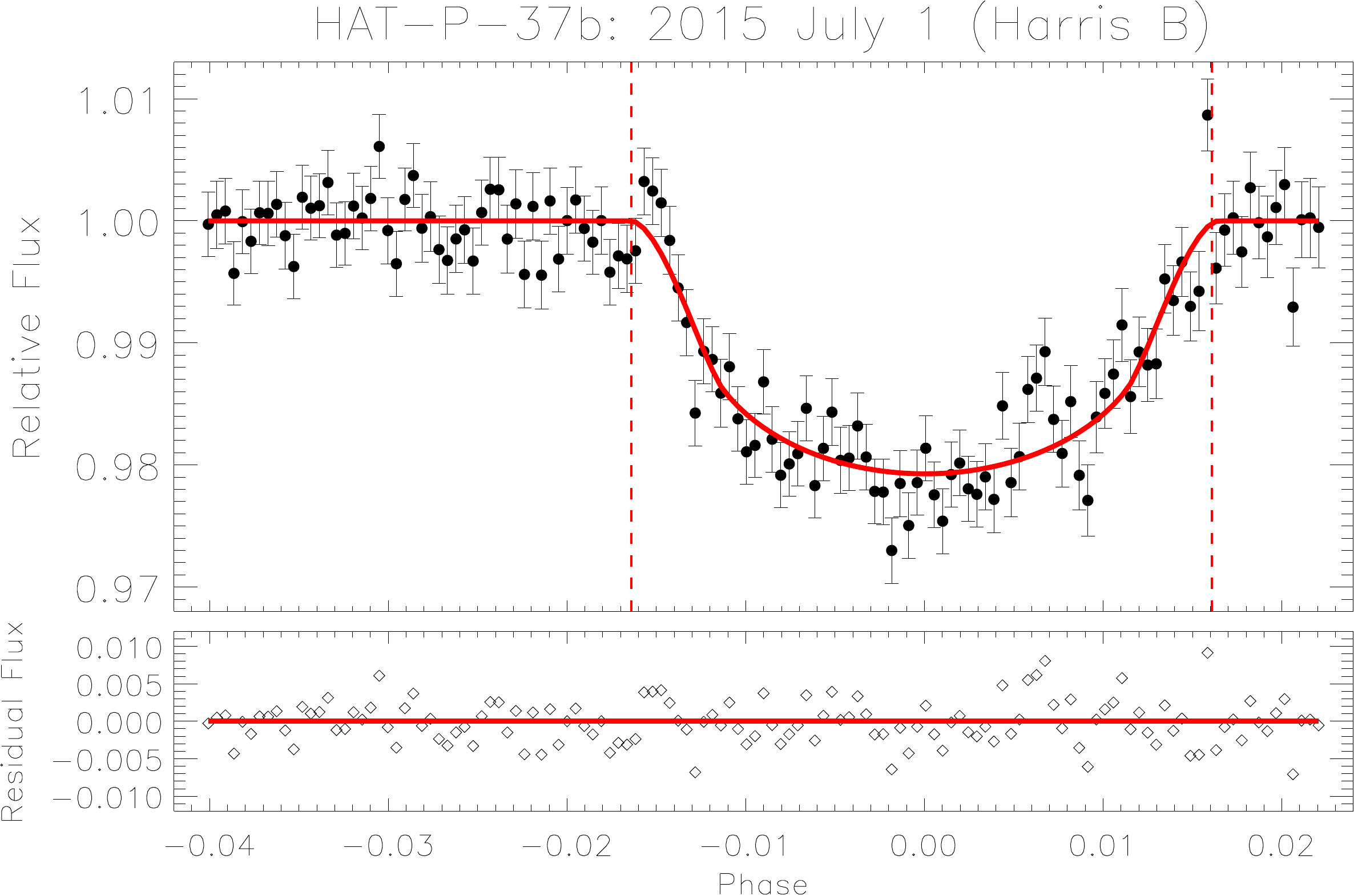} & 
\includegraphics[width=0.33\linewidth,angle=90]{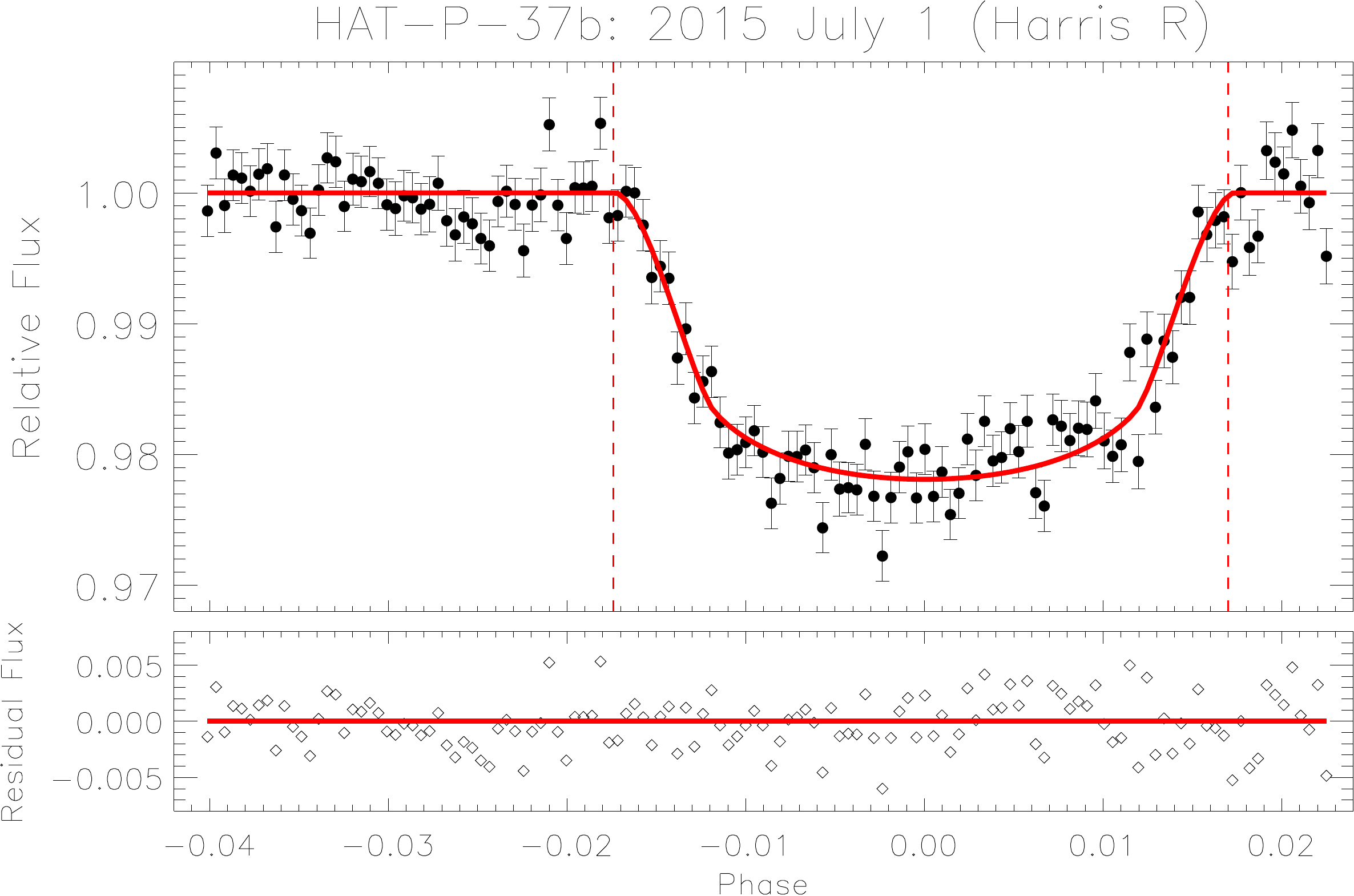} \\ 
\end{tabular}
\caption{Light curves of CoRoT-12b, HAT-P-5b, HAT-P-12b, HAT-P-33b, and HAT-P-37b. The 1$\sigma$ error bars include the readout noise, the Poisson noise, and the flat-fielding error. The best-fitting models obtained from the EXOplanet MOdeling Package (\texttt{EXOMOP}) are shown as a solid red line. The model predicted ingress and egress points from \texttt{EXOMOP} are shown as dashed red vertical lines. The residuals (Light Curve - \texttt{EXOMOP} Model) are shown in the second panel. See Table \ref{tb:obs_new} for the cadence, Out-of-Transit root-mean-squared (RMS) flux, and residual RMS flux for each light curve. The data points for all the transits are available in electronic form (see Table \ref{tb:mr}).}
\label{fig:light_1}
\end{figure*}

\begin{figure*}
\centering
\begin{tabular}{cc}
\vspace{0.5cm}
\includegraphics[width=0.33\linewidth,angle=90]{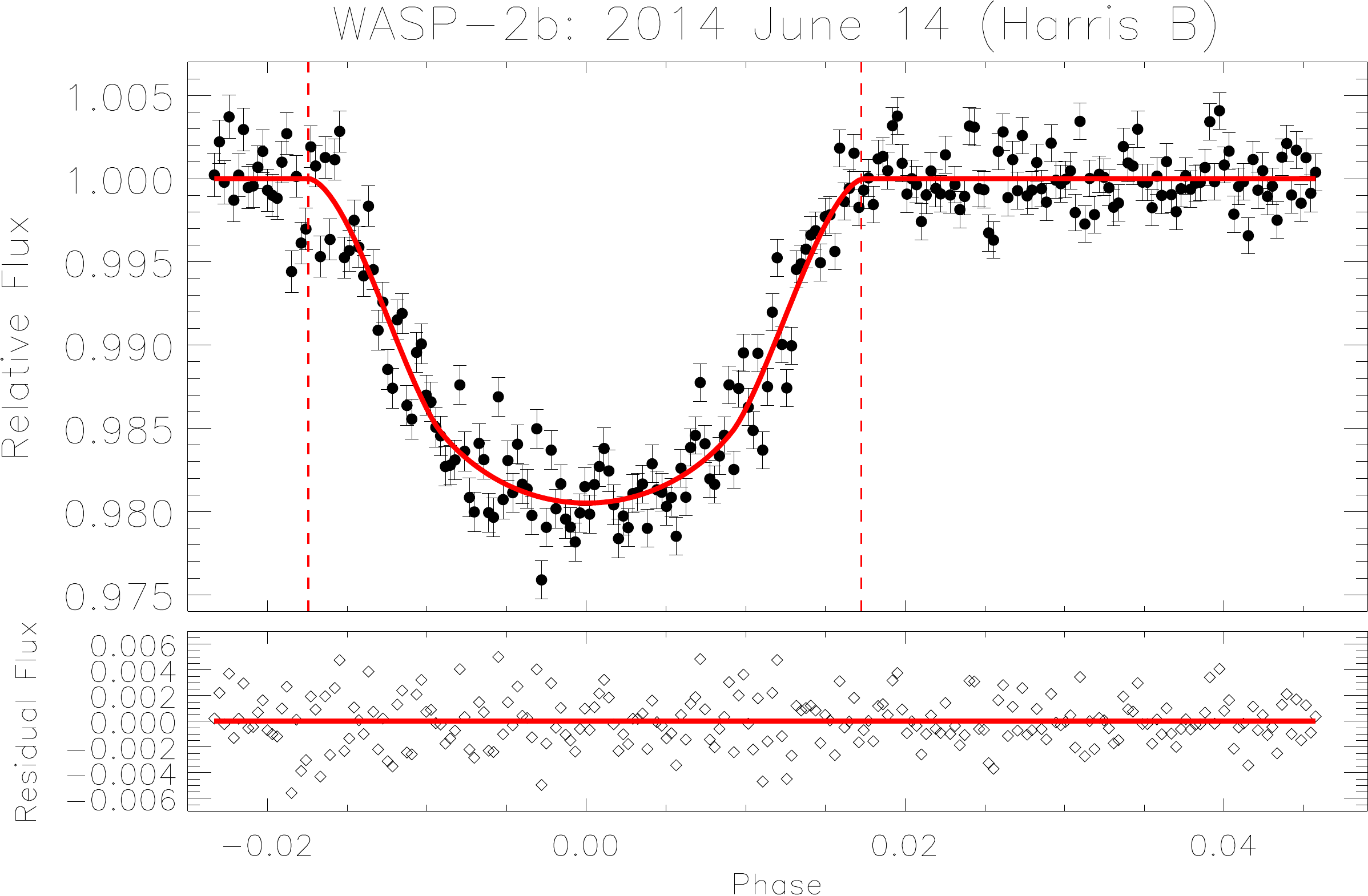}  &
 \includegraphics[width=0.33\linewidth,angle=90]{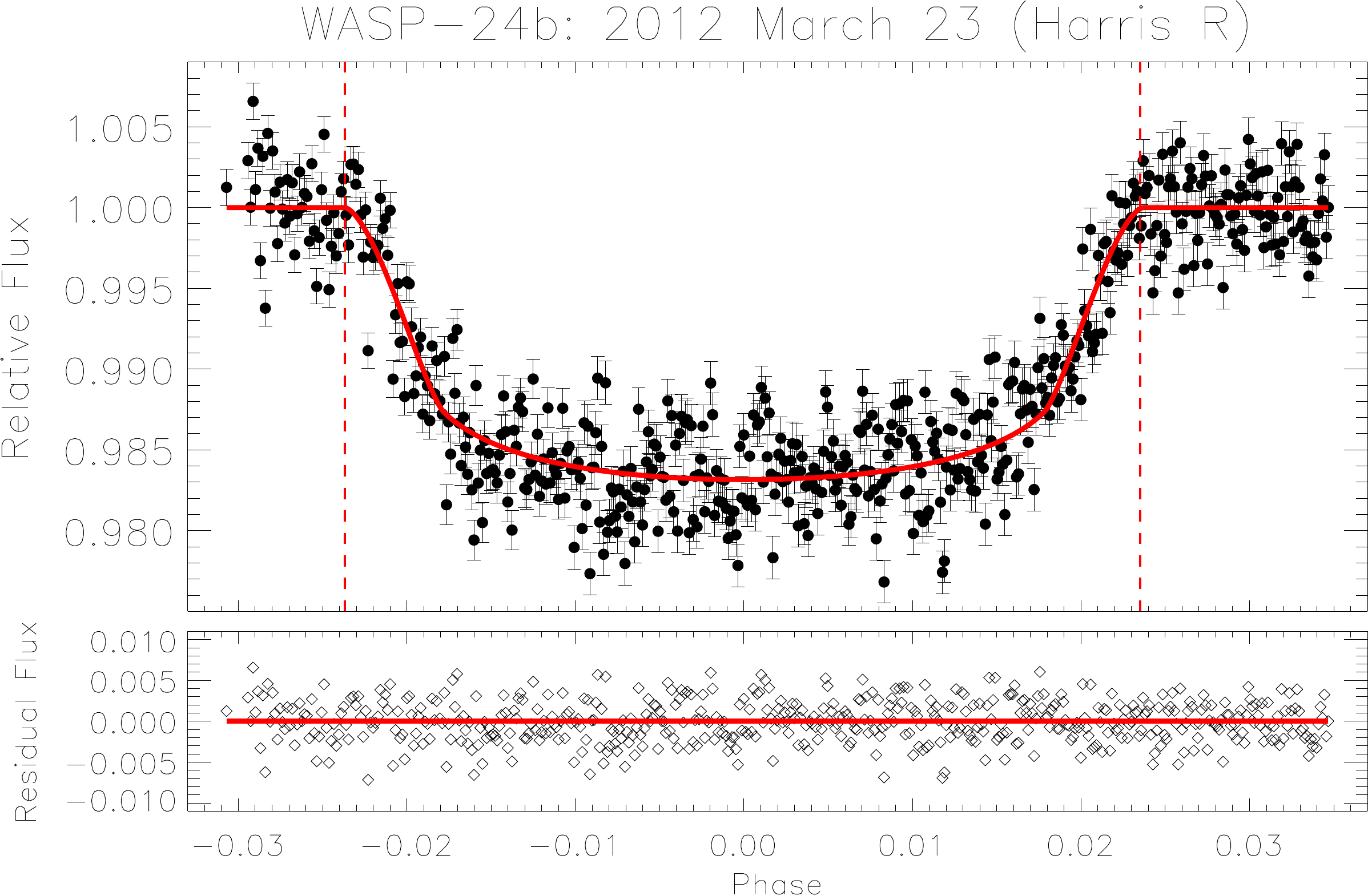} \\
\vspace{0.5cm}
\includegraphics[width=0.33\linewidth,angle=90]{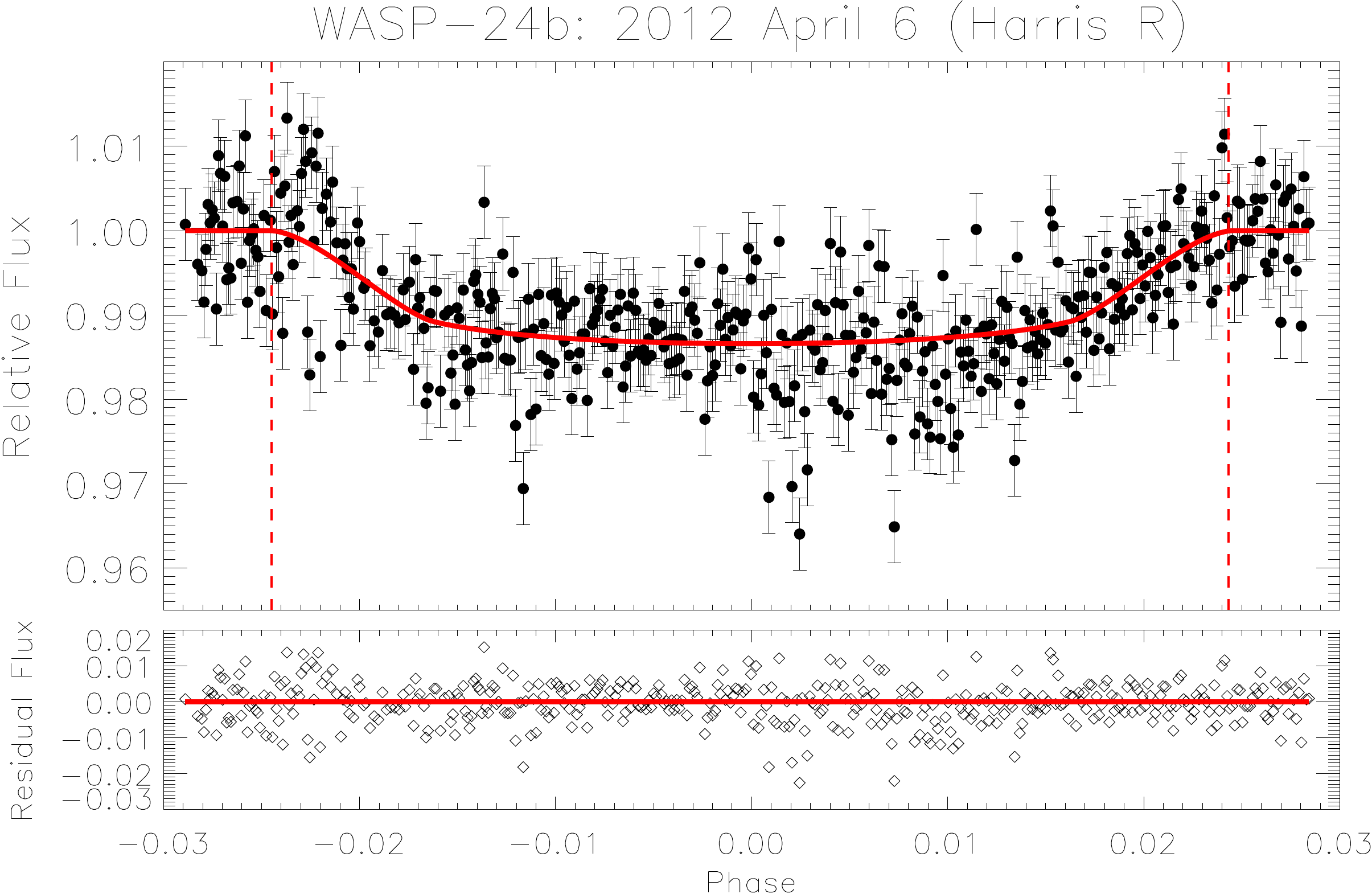} &
\includegraphics[width=0.33\linewidth,angle=90]{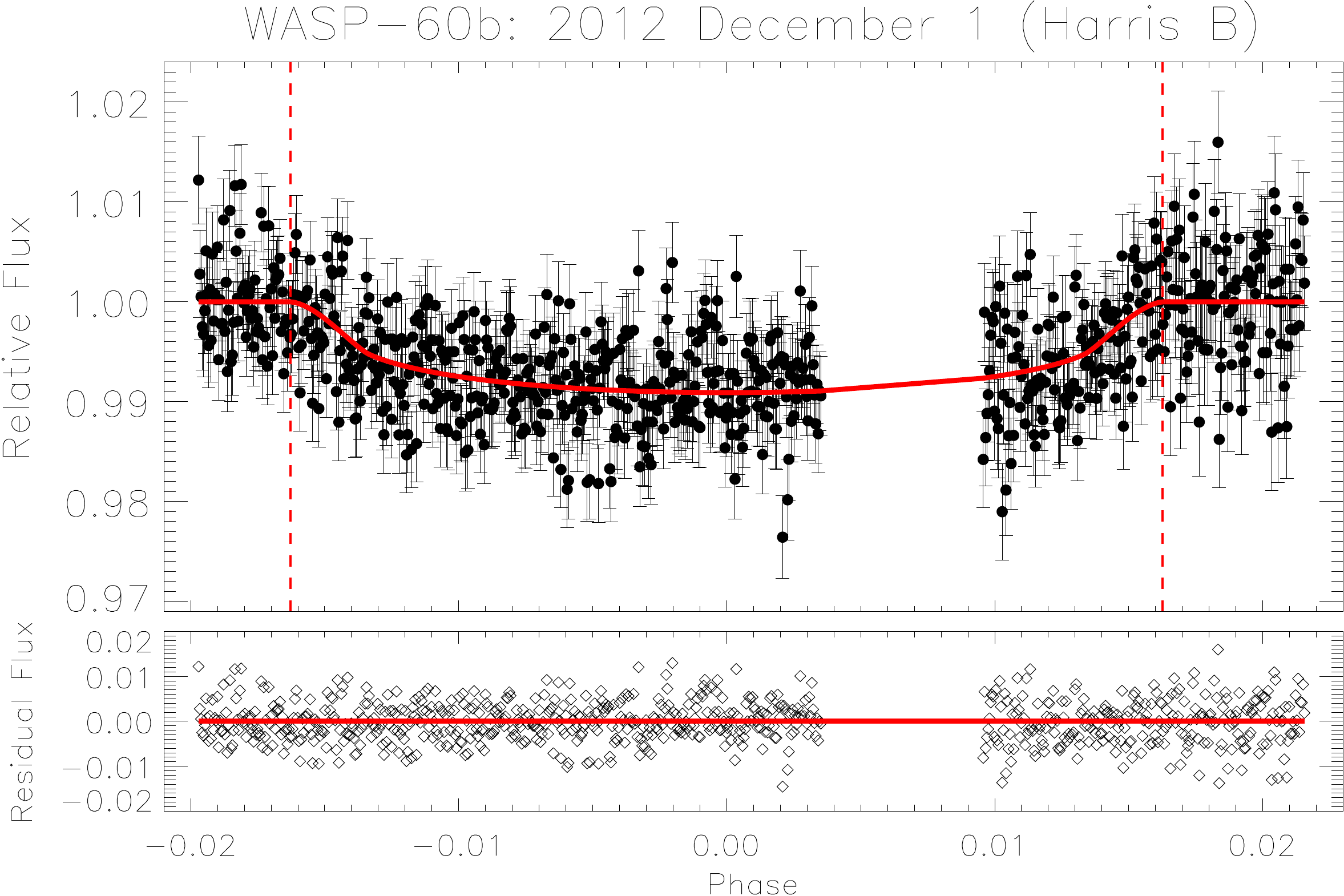} \\
\vspace{0.5cm}
\includegraphics[width=0.33\linewidth,angle=90]{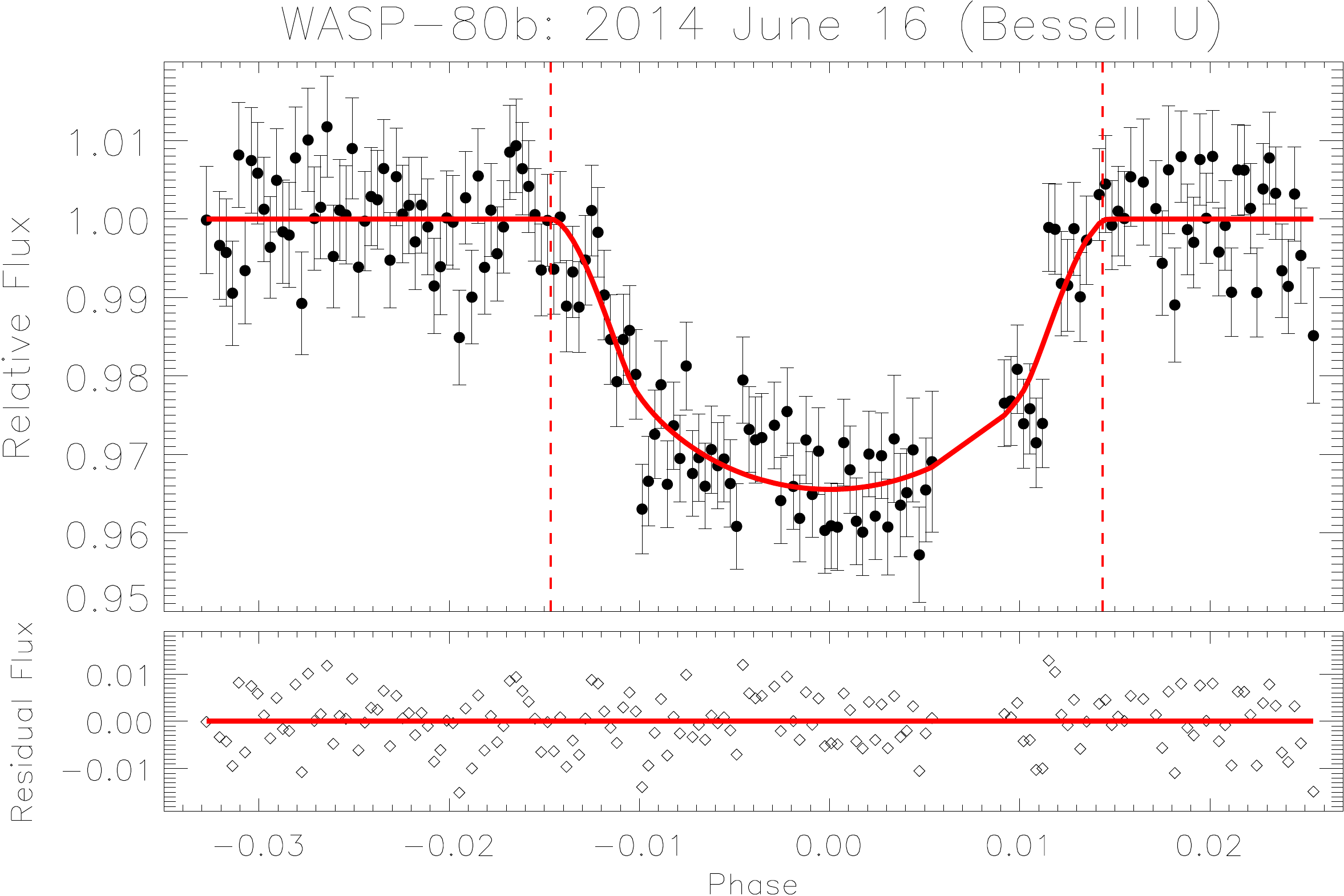} &
\includegraphics[width=0.33\linewidth,angle=90]{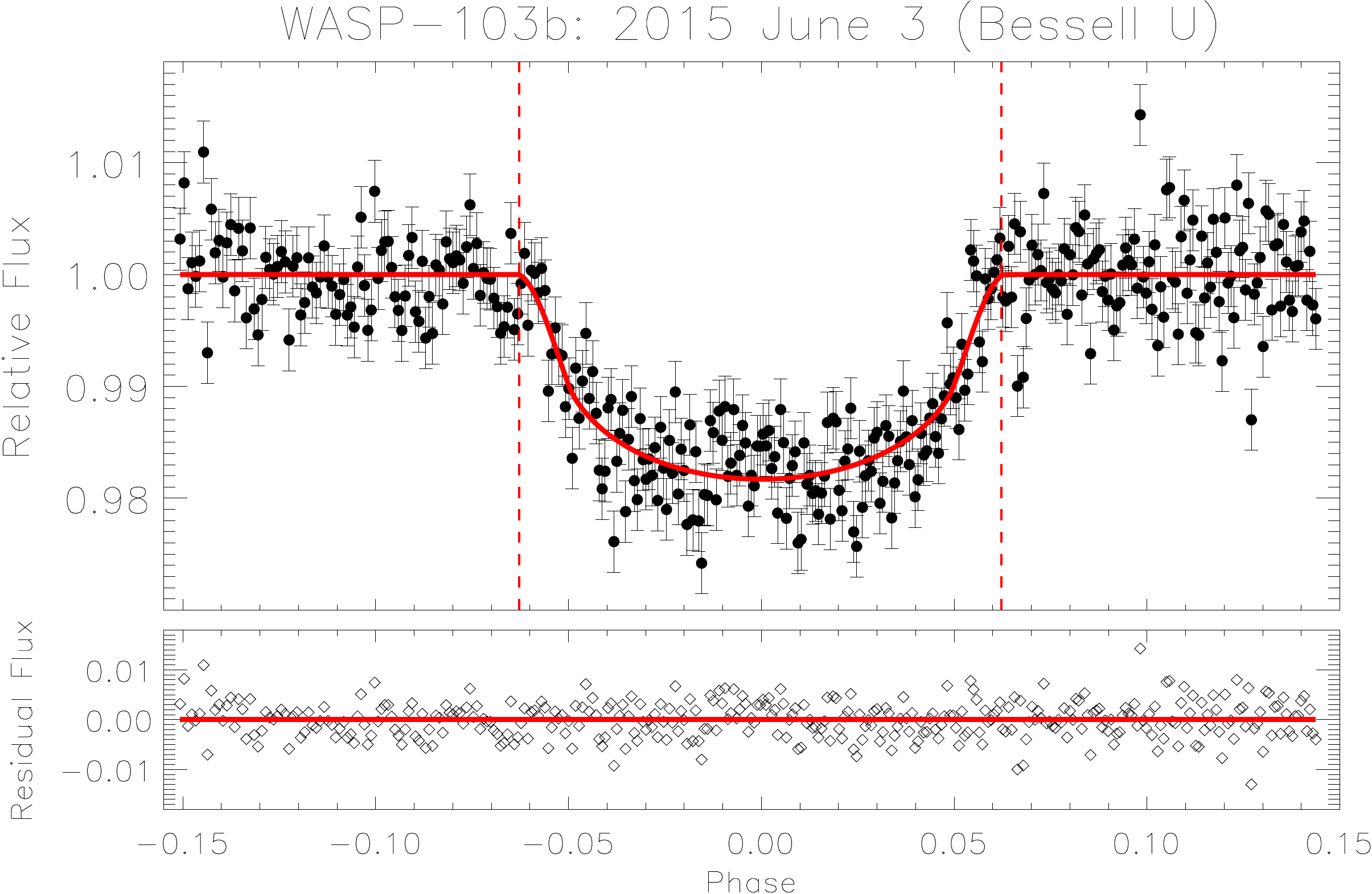} \\
\end{tabular}
\caption{Light curves of WASP-2b, WASP-24b, WASP-60b, WASP-80b, and WASP-103b. Other comments are the same as Fig. 1.}
\label{fig:light_2}
\end{figure*}

\begin{figure*}
\centering
\begin{tabular}{c}
\includegraphics[width=0.37\linewidth,angle=90]{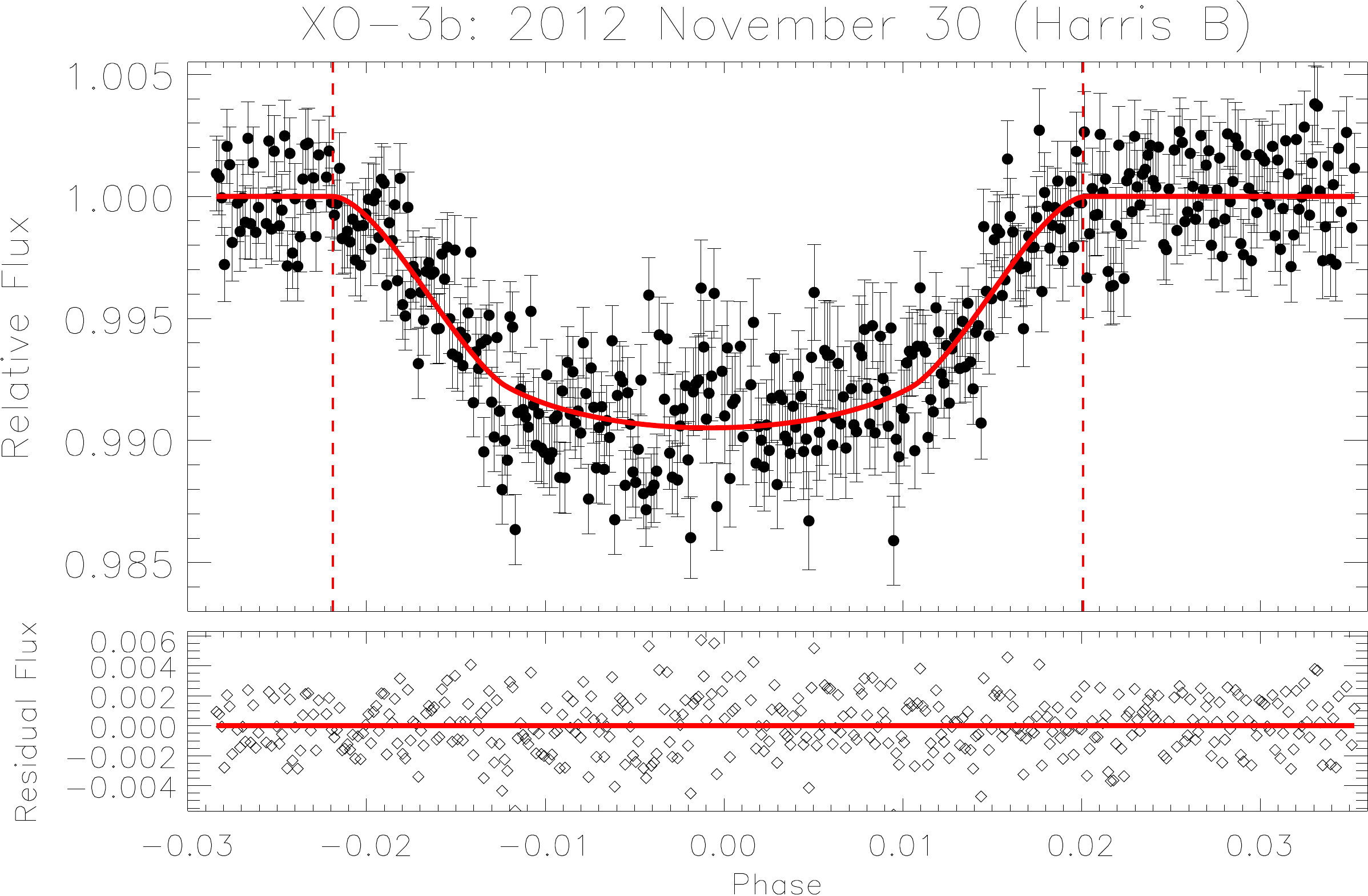} \\
\end{tabular}
\caption{Light curve of XO-3b. Other comments are the same as Fig. 1.}
\label{fig:light_3}
\end{figure*}

To reduce the data and create the light curves we use the reduction pipeline \texttt{ExoDRPL}\footnote{https://sites.google.com/a/email.arizona.edu/kyle-pearson/exodrpl} \citep{Pearson2014}. Each of our images are bias-subtracted and flat-fielded with 10 biases and flats. To produce the light curve for each observation we perform aperture photometry (using \texttt{phot} in the \texttt{\texttt{IRAF}\footnote{\texttt{IRAF} is distributed by the National Optical Astronomy Observatory, which is operated by the Association of Universities for Research in Astronomy, Inc., under cooperative agreement with the National Science Foundation.} DAOPHOT} package) by measuring the flux from our target star as well as the flux from 8 different reference stars with 110 different circular aperture radii. The aperture radii sizes we explore are different for every observation due to changes in seeing conditions. For the analysis, a constant sky annulus for every night of observation of each target is chosen (a different sky annulus is used depending on the seeing and the crowdedness of the target field) to measure the brightness of the sky during the observations. We reduce the risk of contamination by making sure no stray light from the target star or other nearby stars falls in the chosen aperture. A synthetic reference light curve is produced by averaging the light curves from our reference stars. The final light curve of each date is normalized by dividing by this synthetic light curve to correct for any systematic differences from atmospheric variations (i.e. airmass) throughout the night. Every combination of reference stars and aperture radii are considered and we systematically choose the best aperture and reference stars by minimizing the scatter in the Out-of-Transit (OoT) data points. The 1$\sigma$ error bars on the data points include the readout noise, flat-fielding errors, and Poisson noise. The final light curves are presented in Figs. \ref{fig:light_1}--\ref{fig:light_3}. The data points of all our transits are available in electronic form (see Table \ref{tb:mr}). For all the transits, the OoT baselines have a photometric root-mean-squared (RMS) value between 1.13 and 7.76 millimagnitude (mmag).


\begin{table*}
\centering
\caption{Photometry of all our light curves$^{1}$}
\begin{tabular}{cccccccc}
\hline
\hline
 Planet Name           & 
Filter            & 
 Time (BJD$_{TDB}$)           & 
Relative flux  &
Error bars &
CCD X-Pos &
CCD Y-Pos &
Median Airmass \\
\hline
\hline
CoRoT-12b & Harris-R & 2456338.584736 & 0.99648 & 0.003327 & 529.411 & 635.689 & 1.513072 \\
CoRoT-12b & Harris-R & 2456338.585434 & 0.998034 & 0.003213 & 527.611 & 634.955 & 1.507972 \\
CoRoT-12b & Harris-R & 2456338.586133 & 0.998494 & 0.003108 & 528.101 & 634.46 & 1.502944 \\
CoRoT-12b & Harris-R & 2456338.586831 & 0.99549 & 0.003094 & 528.858 & 634.639 & 1.497979 \\
\hline
\end{tabular}
\vspace{-2em}
\tablenotetext{1}{This table is available in its entirety in machine-readable form in the online journal.  A portion is shown here for guidance regarding its form and content. }	
\label{tb:mr}	
\end{table*}

\section{Light Curve Analysis}\label{sec:exomop}
To find the best-fit to the light curves we use the EXOplanet MOdeling Package (\texttt{EXOMOP}; \citealt{Pearson2014}; \citealt{Turner2016})\footnote{\texttt{EXOMOPv7.0} is used in the analysis and is available on Github at \href{https://github.com/astrojake/EXOMOP}{https://github.com/astrojake/EXOMOP} }, which utilizes the analytic equations of \cite{Mandel2002} to generate a model transit. For a complete description of \texttt{EXOMOP} see \citet{Pearson2014} and \citet{Turner2016}. The $\chi^2$-fitting statistic for the model light curve used in \texttt{EXOMOP} is: 
\begin{equation}
   \chi^2 = \sum_{i=1}^{N_{pts}} \left[  \frac{f_{i}( \mbox{obs} )  - f_{i}(\mbox{model})}{\sigma_{i}(\mbox{obs})}    \right]^2      \label{eq:chi22}
\end{equation}
where $N_{pts}$ is the total number of data points (Table \ref{tb:obs_new}), $f_{i}(\mbox{obs})$ is the observed flux at time $i$, $\sigma_{i}(\mbox{obs})$ is the error in the observed flux, and $f_{i}(\mbox{model})$ is the calculated model flux.

\texttt{EXOMOP} uses the following procedure to find a best-fit to the data. A Levenberg-Marquardt (LM) non-linear least squares minimization (\texttt{MPFIT}; \citealt{Markwardt2009}; \citealt{Press1992}) is performed on the data and a bootstrap Monte Carlo technique (\citealt{Press1992}) is used to calculate robust errors of the LM fitted parameters. Additionally, a Differential Evolution Markov Chain Monte Carlo (DE-MCMC; \citealt{terBrrak2002}; \citealt{Eastman2013}) analysis is used to model the data. The fitted parameters that have the highest error bars from either the LM or DE-MCMC best-fitting model are used in the analysis. In every case both models find results within 1$\sigma$ of each other. Additionally, \texttt{EXOMOP} uses the residual permutation (rosary bead; \citealt{Southworth2008}), time-averaging (\citealt{Pont2006}), and wavelet (\citealt{Carter2009}) methods to assess the importance of red noise in both fitting methods. Not accounting for red noise in the data underestimates the fitted parameters (\citealt{Pont2006}; \citealt{Carter2009}). In order to be conservative, the red noise method that produces the largest errors is used to inflate the errors in the fitted parameters. Finally, in order to compensate for underestimated observational errors we multiply the error bars of the fitted parameters by $\sqrt{\chi^{2}_{r}}$ when the reduced chi-squared ($\chi^{2}_{r}$) of the data (Table \ref{tb:obs_new}) is greater than unity (e.g. \citealt{Bruntt2006}; \citealt{Southworth2007a};  \citealt{Southworth2007b};  \citealt{Southworth2008}; \citealt{Barnes2013}; \citealt{Turner2016}). 

\texttt{EXOMOP} uses the Bayesian Information Criterion (BIC; \citealt{schwarz1978}) to assess over-fitting of the data. The BIC is defined as 
\begin{equation}
BIC =  \chi^{2} + k \ln{(N_{pts})}, \label{eq:BIC}
\end{equation}
where $\chi^{2}$ is calculated for the best-fitting model (equation \ref{eq:chi22}) and $k$ is the number of free parameters (Table \ref{tb:obs_new}) in the model fit [$f_{i}(\mbox{model})$]. The power of the BIC is the penalty for a higher number of fitted model parameters, making it a robust way to compare different best-fit models. The preferred model is the one that produces the lowest BIC value. 

Each transit is modeled with \texttt{EXOMOP} using 10000 iterations for the LM model and 20 chains and 20$^{6}$ links for the DE-MCMC model. The Gelman-Rubin statistic (\citealt{Gelman1992}) is used to ensure chain convergence (\citealt{Ford2006}) in the MCMC model. During the analysis of each transit the mid-transit time ($T_{c}$), planet-to-star radius ($R_{p }/R_{*}$), scaled semi-major axis ($a/R_{*}$), and inclination ($i$) are set as free parameters. The previously published values for $a/R_{*}$, $i$, and $a/R_{*}$ are used as priors for the LM model (Table \ref{tb:fit_pars}). The results of the LM fit are used as the prior for the DE-MCMC. The eccentricity ($e$), argument of periastron ($\omega$), and period ($P_{p}$) of each of the planets are fixed (see Table \ref{tb:fit_pars} for their values) in the analysis because these parameters have minimal effect on the overall shape of the light curve. The linear and quadratic limb darkening coefficients in each filter are taken from \citet{Claret2011} and interpolated to the stellar parameters of the host stars (see Table \ref{tb:fit_pars_limb}) using the \texttt{EXOFAST} applet\footnote{http://astroutils.astronomy.ohio-state.edu/exofast/limbdark.shtml}\citep{Eastman2013}. In addition, a linear or quadratic least squares fit is modeled to the OoT baseline simultaneously with the \cite{Mandel2002} model. The BIC is used to determine whether to include any baseline fit in the best-fit model and the baseline with the lowest BIC value is always chosen.


The light curve parameters obtained from the \texttt{EXOMOP} analysis and the derived transit durations are summarized in Table \ref{tb:1_light}. The modeled light curves can be found in Figs. \ref{fig:light_1}--\ref{fig:light_3} and the physical parameters for our targets are derived as outlined in Section \ref{sec:physical_properites} (Tables \ref{tb:pars_1_light}--\ref{tb:pars_2_light}). A thorough description of the modeling and results of each system can be found in Section \ref{sec:indiv_systems}. 

\begin{table*}
\centering
\caption{Parameters fixed for the light curve fitting using \texttt{EXOMOP}}
\begin{tabular}{ccccccc}
  \hline
  \hline
Planet 	&
Period ($P_{p}$) &
a/$R_{\ast}^a$	&
Inclination ($i$)$^a$ & 
Eccentricity ($e$) &
Omega ($\omega$) &
Source  \\
& (days)&     &($^\circ$) &   &($^\circ$)  &    \\
  \hline
  \hline
CoRoT-12b & 2.828042 & 7.7402 & 85.48& 0.070 & 105 & 1 \\
HAT-P-5b & 2.788491 & 7.5 & 86.75 & 0 & 0 & 2\\
HAT-P-12b & 3.213089 & 11.7371 & 89.915  & 0 & 0 & 3 \\
HAT-P-33b & 3.474474 & 6.56 & 87.2  & 0.148 & 96 & 4 \\
HAT-P-37b & 2.797436 & 9.32 & 86.9 & 0.058 & 164 & 5 \\
WASP-2b & 2.15221812 & 8.06 & 84.89 & 0 & 0 & 6 \\
WASP-24b & 2.341213 & 5.98  & 83.64 & 0 & 0 & 7 \\
WASP-60b & 4.3050011 & 10 & 87.9 & 0 & 0 & 8 \\
WASP-80b & 3.06785 & 12.989 & 89.92 & 0.07 & 0 & 9 \\
WASP-103b & 0.925542 & 2.978 & 86.3 & 0 & 0 & 10 \\
XO-3b & 3.191524 & 7.07  & 84.2 & 0.26 & 345.8 & 11 \\
  \hline
\end{tabular}
\vspace{-2em}
\tablenotetext{a}{These parameter values were not fixed in the final analysis but were used as priors for the MCMC.}
\tablerefs{(1) \citealt{Gillon2010}; 
(2) \citealt{Bakos2007};
(3) \citealt{Lee2012}; 
(4new) \citealt{Hartman2011};
(5) \citealt{Bakos2012};
(6) \citealt{Becker2013};
(7) \citealt{Street2010};
(8) \citealt{Hebrard2013};
(9) \citealt{Triaud2013};
(10) \citealt{Gillon2014};
(11) \citealt{Winn2008}
}
\label{tb:fit_pars}
\end{table*}

\begin{table*}
\centering
\caption{Limb darkening coefficients for the light curve fitting using \texttt{EXOMOP}}
\begin{tabular}{cccccccc}
\hline
\hline
Planet	&
Filter           & 
Linear coefficient$^{1}$ &
Quadratic coefficient$^{1}$   &
T$_{\textrm{eff}}$ [K]&
[Fe/H]	&
$\log{g}$ [cgs] &
Source \\
\hline   
\hline
CoRoT-12b 	& R 	& 0.39440901 & 0.26682249	& 5675	& 0.160 & 4.375 & 1 \\
HAT-P-5b & U & 0.75552025 & 0.093484300 & 5960 & 0.240 & 4.368          & 2 \\
HAT-P-12b & B & 0.93774724 & -0.083432883 & 4650 & -0.290 & 4.610       & 3 \\
HAT-P-33b & R &0.27628872 &0.32295169 &6401 & 0.05 & 4.15               & 4\\
HAT-P-37b & B & 0.72832760 & 0.097543998 & 5500 & 0.03 & 4.52           & 5\\
HAT-P-37b & R & 0.41967640 & 0.25020840 & 5500 & 0.03 & 4.52            & 5\\
WASP-2b & B & 0.82272126 & 0.018632333 & 5200 & 0.100 & 4.537           & 2\\
WASP-24b & R & 0.31410756 & 0.30624587 & 6080 & -0.002 & 4.26           & 6 \\
WASP-60b & B & 0.61455358 & 0.18477201 & 5900 & -0.040 & 4.20           & 7 \\
WASP-80b & U & 0.82663825 & -0.029831771 & 4150 & -0.140 & 4.60         &8\\
   WASP-103b & U & 0.65536932 & 0.17875591 & 6110 & 0.060 & 4.22        &9 \\
   XO-3b & B & 0.50449954 & 0.25897027 & 6429 & -0.177 & 3.950          &10 \\
\hline   
\end{tabular}
\vspace{-2em}
\tablenotetext{1}{The limb darkening coefficients are taken from \citet{Claret2011} and interpolated to the stellar parameters of their host star using \texttt{EXOFAST}}
\tablerefs{
(1) \citealt{Gillon2010}; 
(2) \citealt{Torres2008};
(3) \citealt{Hartman2009};
(4)  \citealt{Hartman2011};
(5)  \citealt{Bakos2012};
(6) \citealt{Street2010}; 
(7) \citealt{Hebrard2013};
(8) \citealt{Triaud2013}; 
(9); \citealt{Gillon2014}; 
(10) \citealt{Johns-Krull2008};
}
\label{tb:fit_pars_limb}
\end{table*}

\begin{table*}
\centering
\caption{Light curve parameters derived in this study using \texttt{EXOMOP}}
\begin{tabular}{cccccc}
\hline
\hline
Planet                  &CoRoT-12b                      &HAT-P-5b               &HAT-P-12b                      &HAT-P-33b              & HAT-P-37b    \\
Date                    &2013 February 15               &2015 June 6            &2014 January 19                &2012 April 6           &2015 July 1 \\
Filter$^{1}$            &R                              &U                      &B                              &R                      &B\\
$T_{c}$ (BJD$_{TDB}$-2450000)   &6338.67097$\pm0.00074$ &7180.82658$\pm0.00076$	&6677.97482$\pm0.00047$  	    &6024.71746$\pm0.0012$   &7205.91376$\pm0.00054$ \\
R$_p$/R$_\ast$          &0.1645$^{+0.0038}_{-0.0040}$   &0.1225$\pm0.0051$      & 0.1386$^{+0.0013}_{-0.0014}$  & 0.1152$\pm0.0017$     & 0.1253$\pm0.0021$   \\
$a/R_{*}$               &6.59$^{+0.31}_{-0.29}$         &6.05$\pm0.44$          & 11.86$\pm0.57$                &5.67$\pm0.13$          &10.82$\pm0.91$\\
Inclination ($^\circ$)  &83.54$\pm0.71$                 &83.31$\pm1.11$         & 90.98$\pm1.09$                & 90.08$\pm3.43$        &89.99$1.83$ \\
Duration (mins)         &174.0$\pm1.4$                  &184.1$\pm1.9$          & 139.8$\pm4.4$                 & 270.45$\pm0.48$       &132.8 $\pm$ 2.7\\
Red noise (mmag)        & 0.0001                        &1.60                   &  0.21                         &0.78                   & 0.0001\\
OoT baseline function   & None                          &None                   & Quadratic                     &None                   &None\\
\hline
Planet                  &HAT-P-37b                      &HAT-P-37b          &WASP-2b                &WASP-24b       &WASP-24b\\
Date                    &2015 July 1                    &Weighted Average   &2014 June 14           &2012 March 23  &2012 April 6                      \\
Filter$^{1}$            &R                              &---                &B                      &R                  &R                       \\
$T_{c}$ (BJD$_{TDB}$-2450000)   &7205.91325$\pm0.00056$         &---                &6823.83839$\pm0.00055$  &6010.8437$-0.0017$ & 6024.8910$\pm0.0015$                      \\
R$_p$/R$_\ast$          &0.1361$\pm0.0028$              &0.1291$\pm0.0017$  &0.1383$\pm$0.0049      &	0.1139$\pm$0.0015 &  	0.1113$\pm0.0043$                     \\
$a/R_{*}$               &9.14$\pm0.63$                  &9.68$\pm0.52$      &8.05$\pm{1.21}$	    &  7.42$\pm0.15$        &    6.06$\pm0.73$                 \\
Inclination ($^\circ$)  &86.73$\pm0.93$                 &87.4$\pm0.82$      &84.86$\pm1.61$         &   90.0$\pm5.4$    &83.95$\pm2.74$                  \\
Duration (mins)         &140.5 $\pm$ 2.7                &136.6$\pm$1.9      &108.4$\pm1.4$          &  159.5$\pm0.6$    &  165.0$\pm0.6$                   \\
Red noise (mmag)        &0.34                           &---                &0.00                    & 0.27       &0.60                     \\
OoT baseline function   &None                           &---                &Quadratic             &Quadratic              &Linear                    \\
\hline
Planet                  &WASP-24b           & WASP-60b              &WASP-80b               &WASP-103b      &XO-3b\\
Date                    & Weighted Average  &2012 December 1        &2014 June 16           &2015 June 3    &2012 November 30\\
Filter$^{1}$            & R                 &B                      &U                      &U              &B\\
$T_{c}$ (BJD$_{TDB}$-2450000)   &---                &6263.6330$\pm0.0012$   &6824.88661$\pm0.00091$ &7177.8222$^{+0.0015}_{-0.0009}$    
&6262.6566$\pm0.0015$\\
R$_p$/R$_\ast$          &0.1136$\pm0.0014$  &	0.0852$\pm0.0036$   &0.1615$\pm0.0033$      &0.1181$\pm$0.0016 &0.0968$\pm0.0023$\\
$a/R_{*}$               &7.36$\pm$0.15      &9.49$\pm1.81$          &12.85$\pm0.42$         &2.90$\pm0.05$&	5.68$\pm0.51$\\
Inclination ($^\circ$)  &85.19$\pm2.44$     &	87.48$\pm2.83$      &90.0$\pm1.8$           &	90.00$\pm0.18$ &81.75$\pm0.77$	\\
Duration (mins)         &162.1$\pm0.4$      &	201.9$\pm0.3$       &126.7 $\pm2.2$         &167.6 $\pm1.5$&	185.4 $\pm0.9$\\
Red noise (mmag)        &---                &0.00                   &0.01                   &0.001  &0.001\\
OoT baseline function   &---                &None                   &Quadratic               &Linear&None\\
\hline
 \end{tabular}
 \vspace{-2em}
\tablenotetext{1}{Filter: U is the Bessell U (303--417 nm), B is the Harris B (330--550 nm), and R is the Harris R (550--900 nm)  }
\label{tb:1_light}
\end{table*}

\section{Physical Properties of the Systems}\label{sec:physical_properites}
We use the results of our light curve modeling with \texttt{EXOMOP} combined with other measurements in the literature to calculate the planetary mass (e.g. \citealt{Winn2010b}; \citealt{Seager2011}), radius, density, surface gravity (e.g. \citealt{Southworth2007a}), modified equilibrium temperature (e.g. \citealt{Southworth2010a}), Safronov number (e.g. \citealt{Safronov1972}; \citealt{Southworth2010a}), and atmospheric scale height (e.g. \citealt{Seager2011}; \citealt{deWit2013}). An updated period and ephemeris is also calculated and is described in detail in Section \ref{sec:period}. To calculate the physical parameters we use the values from the modeling ($P_{p}$, $R_{p}/R_{\ast}$, $i$, $a/R_{\ast}$) and for the orbital ($e$) and host star parameters (radial velocity amplitude, mass, radius, equilibrium temperature) we use the values found in the literature. When calculating the scale height, the mean molecular weight in the planet's atmosphere was set to 2.3 assuming a H/He-dominated atmosphere (\citealt{deWit2013}). The physical parameters of all our systems can be found in Tables \ref{tb:pars_1_light}--\ref{tb:pars_2_light}.

 \begin{table*}
\centering
\caption{Physical parameters derived in this study for CoRoT-12b, HAT-P-5b, HAT-P-12b,HAT-P-33b, HAT-P-37b, WASP-2b, WASP-24b, WASP-60b, and WASP-80b. }
\begin{tabular}{cccc}
\hline
\hline

Planet & CoRoT-12b & HAT-P-5b & HAT-P-12b  \\
\hline

Date & 2013 February 15 &  2015 June 6  & 2014 January 19 
\\
Period (Days)
&2.828051 $\pm0.000080$ 
& 2.78847280 $\pm0.00000039$ 
& 3.21305761 $\pm0.00000020$
  \\
$T_{c}(0)$	(BJD-2450000)		
&4398.628 $\pm0.055$ 
& 4241.77716 $\pm0.00015$ 
& 4187.85623 $\pm0.00013$ 
\\     
M$_{b}$ (M$_{Jup}$)  		     
&0.922 $\pm$ 0.072
& 1.06 $\pm$ 0.12  
& 0.211$\pm$0.012 
\\

Our R$_{b}$ (R$_{Jup}$)	
&1.79 $\pm$ 0.15 
& 1.36 $\pm$ 0.057 
& 0.949$\pm$0.017
\\

Reference R$_{b}$ (R$_{Jup}$)	
&1.44 $\pm$ 0.13 (a) 
&1.26 $\pm$ 0.05 (b)
&0.959$\pm$0.029 (c)
\\

$\rho_{b}$ (cgs)	
&0.200 $\pm$ 0.054 
& 0.531 $\pm$ 0.088 
& 0.306$\pm$0.023
\\	

$\log{g_{b}}$ (cgs)			
&2.72 $\pm$ 0.12 
&2.946 $\pm$ 0.085
& 2.77$\pm$0.052 
\\

T$^{'}_{eq}$ (K)			
&1563 $\pm$ 22 
& 1713 $\pm$ 29
& 954$\pm$12
 \\
 
$\Theta$				    
&0.0327 $\pm$ 0.0054 
& 0.0431 $\pm$ 0.0064 
& 0.0235$\pm$0.0019
\\
 
a (au)						
&0.0342 $\pm$ 0.0032 
& 0.0320 $\pm$ 0.0023
& 0.0386$\pm$0.0019
\\

H (km)
&1521 $\pm$ 402 
& 979 $\pm$ 193 
& 822 $\pm$ 98 
\\

\hline
Planet & HAT-P-33b & HAT-P-37b  & WASP-2b  \\
\hline

Date & 2012 April 6  & 2015 July 1  & 2014 June 14  \\
Period (Days)
& 3.4744750$\pm$0.00000037
& 2.79744149 $\pm$0.00000083 
& 2.15222114 $\pm$0.00000019      
\\
$T_{c}(0)$ (BJD-2450000)
& 5110.92726$\pm0.00012$
& 5616.96710$\pm$ 0.00028
& 3991.515553  $\pm0.000074$ 
\\
M$_{b}$ (M$_{Jup}$)
& 1.26$\pm$0.23
& 1.17 $\pm$ 0.10 
& 0.880 $\pm$ 0.087 
\\

Our R$_{b}$ (R$_{Jup}$)
& 1.99$\pm$0.32
& 1.16 $\pm$ 0.06
& 1.12 $\pm$ 0.13
\\

Reference R$_{b}$ (R$_{Jup}$)	
& 1.827$\pm$0.290  (d)
&1.178 $\pm$ 0.077 (e)
&1.043 $\pm$ 0.033 (f)
\\

$\rho_{b}$ (cgs)
& 0.20$\pm$0.10
& 0.93 $\pm$ 0.17 
& 0.77 $\pm$ 0.28 
\\	

$\log{g_{b}}$ (cgs)
& 2.83$\pm$0.21
& 3.369 $\pm$ 0.088 
& 3.25 $\pm$ 0.18
\\

T$^{'}_{eq}$ (K)
& 1901$\pm$26
& 1250 $\pm$ 22 
& 1284 $\pm$ 20 
\\

$\Theta$	
& 0.042 $\pm$ 0.013
& 0.085 $\pm$ 0.012 
& 0.058 $\pm$ 0.016 
\\

a (au)
& 0.0468 $\pm$ 0.0075
& 0.0394 $\pm$ 0.0029 
& 0.0312 $\pm$ 0.0056
\\

H (km)
& 1408$\pm$679
&270 $\pm$ 54 
& 364 $\pm$ 155 
\\
\hline
Planet & WASP-24b  & WASP-60b  & WASP-80b
\\
\hline

Date & Combined &  2012 December 1 & 2014 June 16 
\\
Period (Days)
& 2.34121877 $\pm0.00000030$ 
& 4.305022 $\pm0.000021$ 
& 3.06785925 $\pm0.00000047$ 
\\

$T_{c}(0)$ (BJD-2450000)
& 4945.589444 $\pm0.000090$ 
& 5747.0302$\pm0.0022$ 
& 6125.418034$\pm0.000052$ 
\\
M$_{b}$ (M$_{Jup}$)
& 1.032 $\pm$ 0.037 
& 0.512$\pm$ 0.034 
& 0.551$\pm$ 0.036 
\\

Our R$_{b}$ (R$_{Jup}$)	
& 1.27 $\pm$ 0.055 
& 0.94$\pm$ 0.12 
& 0.99$\pm$ 0.24 
\\

Reference R$_{b}$ (R$_{Jup}$)
&1.303 $\pm$ 0.047 (g) 
&0.86$\pm$ 0.12 (h)
&0.999$\pm$ 0.031 (i)
\\

$\rho_{b}$ (cgs)
& 0.628 $\pm$ 0.085
& 0.75$\pm$ 0.27 
& 0.71$\pm$ 0.51 
\\	

$\log{g_{b}}$ (cgs)	
& 3.279 $\pm$ 0.057 
& 3.11$\pm$ 0.22 
& 3.22$\pm$ 0.29 
\\

T$^{'}_{eq}$ (K)
& 1583 $\pm$ 27 
& 1354$\pm$ 23 
& 817$\pm$ 20 
\\

$\Theta$	
& 0.0566 $\pm$ 0.0043
& 0.051$\pm$ 0.013 
& 0.072$\pm$ 0.025 
\\

a (au)
& 0.0392 $\pm$ 0.0018
& 0.050$\pm$ 0.011 
&0.0376 $\pm$ 0.0090 
\\

H (km)
& 421 $\pm$ 55
& 536 $\pm$ 274
& 248 $\pm$ 168 \\
\hline
\end{tabular}
\vspace{-2em}
\tablerefs{
(a) \citealt{Gillon2010}; 
(b) \citealt{Bakos2007};
(c) \citealt{Hartman2009}; 
(d) \citealt{Hartman2011}; 
(e) \citealt{Bakos2012};
(f) \citealt{Southworth2010b};
(g) \citealt{Southworth2014};
(h) \citealt{Hebrard2013};
(i)   \citealt{Triaud2015}
}
\label{tb:pars_1_light}
\end{table*}

 \begin{table*}
\centering
\caption{Physical parameters derived in this study for WASP-103b and XO-3b}
\begin{tabular}{ccc}
\hline
\hline
Planet & WASP-103b & XO-3b  \\
\hline

Date & 2015 June 3 &  2012 November 30  \\

Period (days)
& 0.9255454 $\pm0.0000010$
& 3.19153125 $\pm0.00000053$ 
\\

$T_{c}(0)$ (BJD-2450000)
& 6459.59948 $\pm0.00041$ 
& 2997.72200 $\pm0.00040$ 
\\

M$_{b}$ (M$_{Jup}$)
& 1.484 $\pm$ 0.082
& 13.07$\pm$ 0.66 
\\

Our R$_{b}$ (R$_{Jup}$)	
& 1.640 $\pm$ 0.066
& 1.403$\pm$ 0.093 
\\

Previous R$_{b}$ (R$_{Jup}$)
&1.528 $\pm$ 0.073 (a)
&1.217$\pm$ 0.073 (b) 
\\

$\rho_{b}$ (cgs)
&0.417 $\pm$ 0.055 
& 5.87$\pm$ 1.20
\\

$\log{g_{b}}$ (cgs)	
& 3.114 $\pm$ 0.055
& 4.05$\pm$ 0.11
\\

T$^{'}_{eq}$ (K)
& 2537 $\pm$ 42
& 2011$\pm$ 13 
\\

$\Theta$
& 0.0286 $\pm$ 0.0023
& 0.519$\pm$ 0.075
\\

a (au)
& 0.09361 $\pm$ 0.00078 
& 0.0393$\pm$ 0.0041 
\\

H (km)
& 986 $\pm$ 125 
& 90 $\pm$ 22 
\\
\hline
\end{tabular}
\vspace{-2em}
\label{tb:pars_2_light}
\tablerefs{
(a) \citealt{Gillon2014}; (b) \citealt{Winn2008}
}
\end{table*}

\subsection{Period Determination}\label{sec:period}
By combining our mid-transit times found using \texttt{EXOMOP} with previously published mid-transit times, we can refine the orbital period of the targets. When necessary, the mid-transit times were transformed from HJD, which is based on UTC time, into BJD, which is based on Barycentric Dynamical Time (TDB), using the online converter\footnote{http://astroutils.astronomy.ohio-state.edu/time/hjd2bjd.html} by \citet{Eastman2010}. A refined ephemeris for each target is found by performing a weighted linear least-squares analysis using the following equation:
\vspace{-1em}
\begin{equation}
   T_{c} = T_{c}(0) + P_{p}\times{E}, \label{eq:Tc}	
\end{equation} 
where $T_{c}(0)$ is the mid-transit time at the discovery epoch measured in $BJD_{TDB}$, $P_{p}$ is the orbital period of the target and $E$ is the integer number of cycles after their discovery paper. See Tables \ref{tb:pars_1_light}-\ref{tb:pars_2_light} for an updated T$_{c}$ and P$_{b}$ for each system.

\begin{figure*}
\centering
\begin{tabular}{cc}
\vspace{0.5cm}
\includegraphics[width=0.37\linewidth,angle=90]{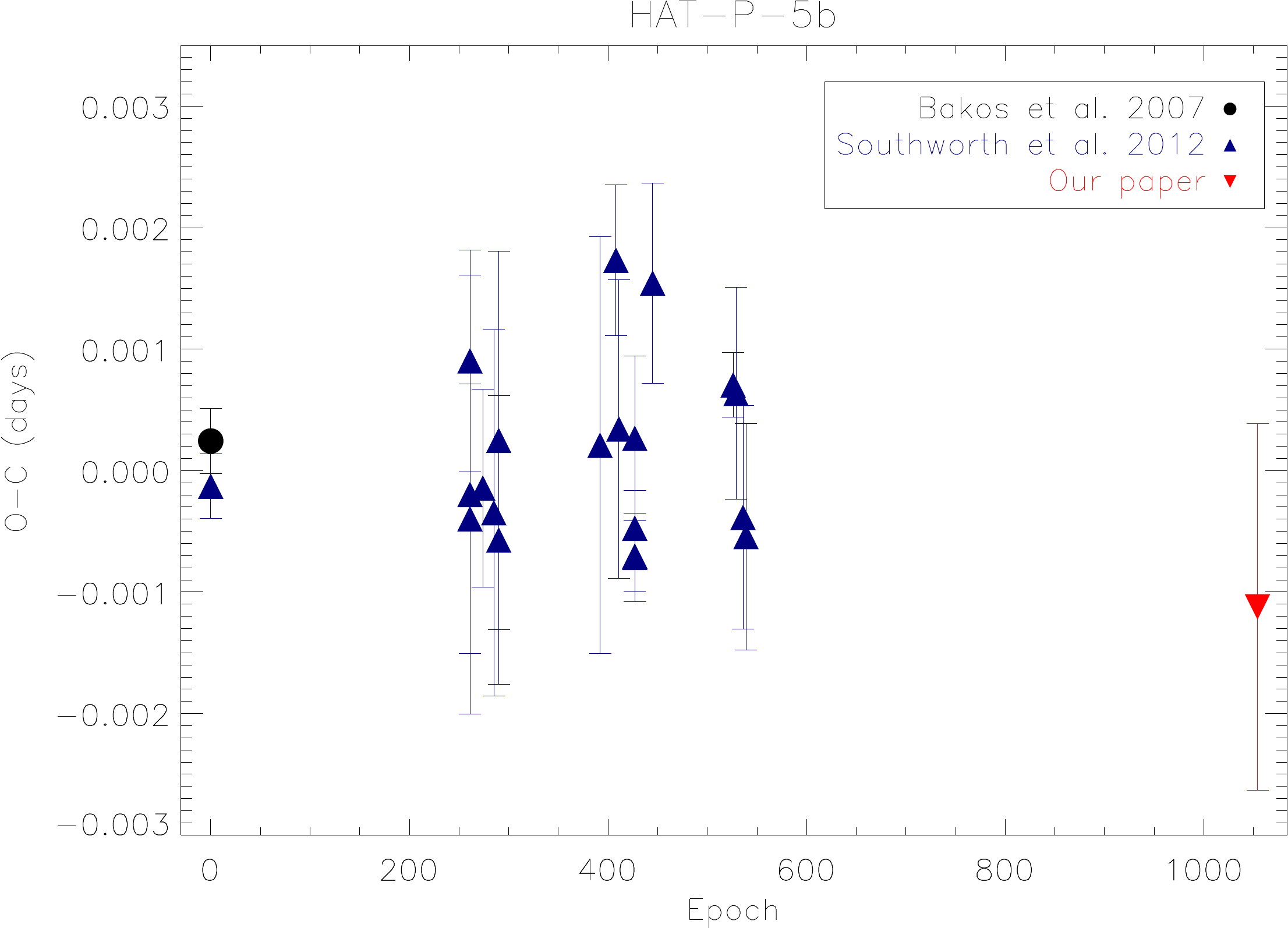} &
\includegraphics[width=0.37\linewidth,angle=90]{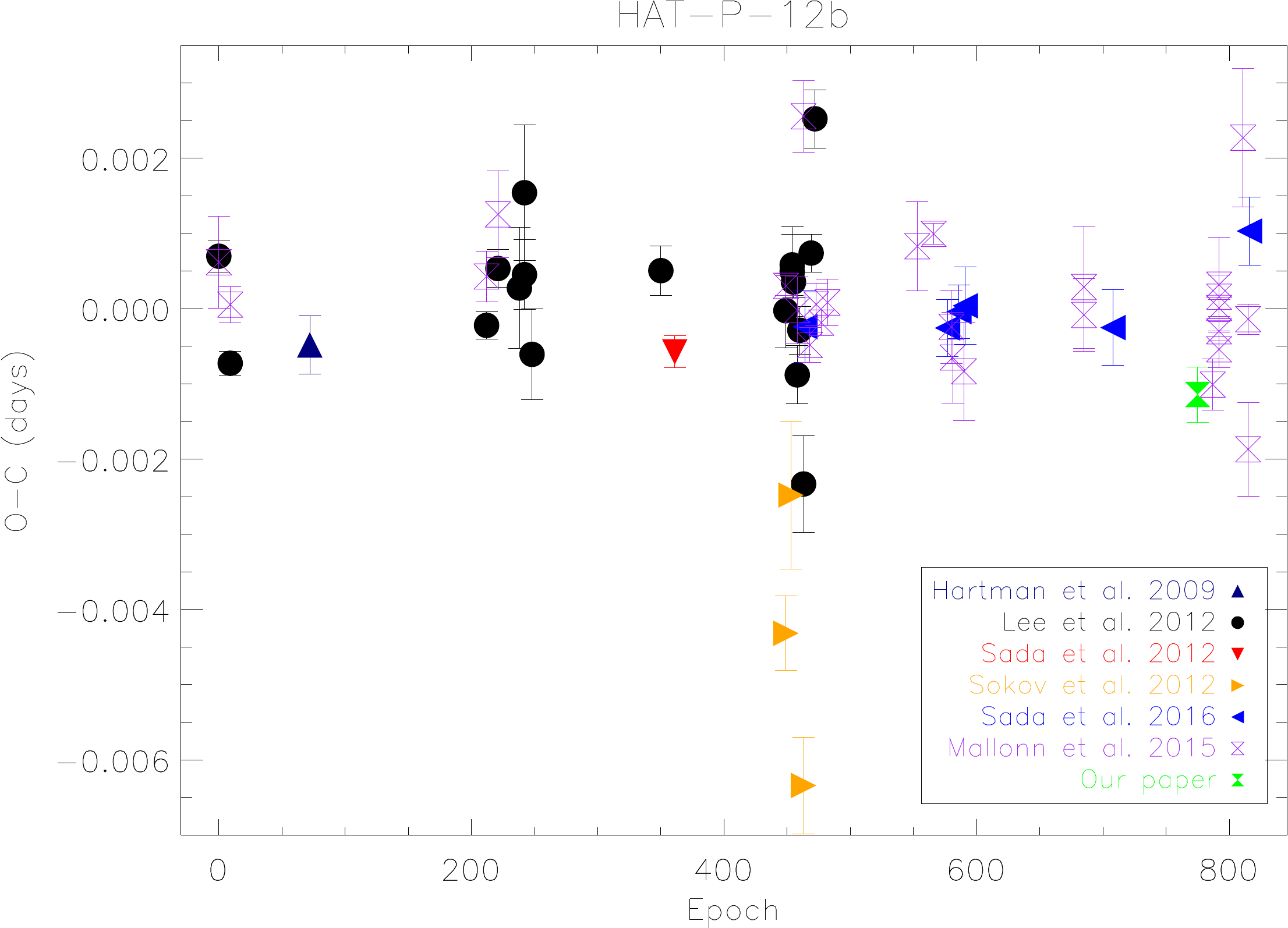} \\
\includegraphics[width=0.37\linewidth,angle=90]{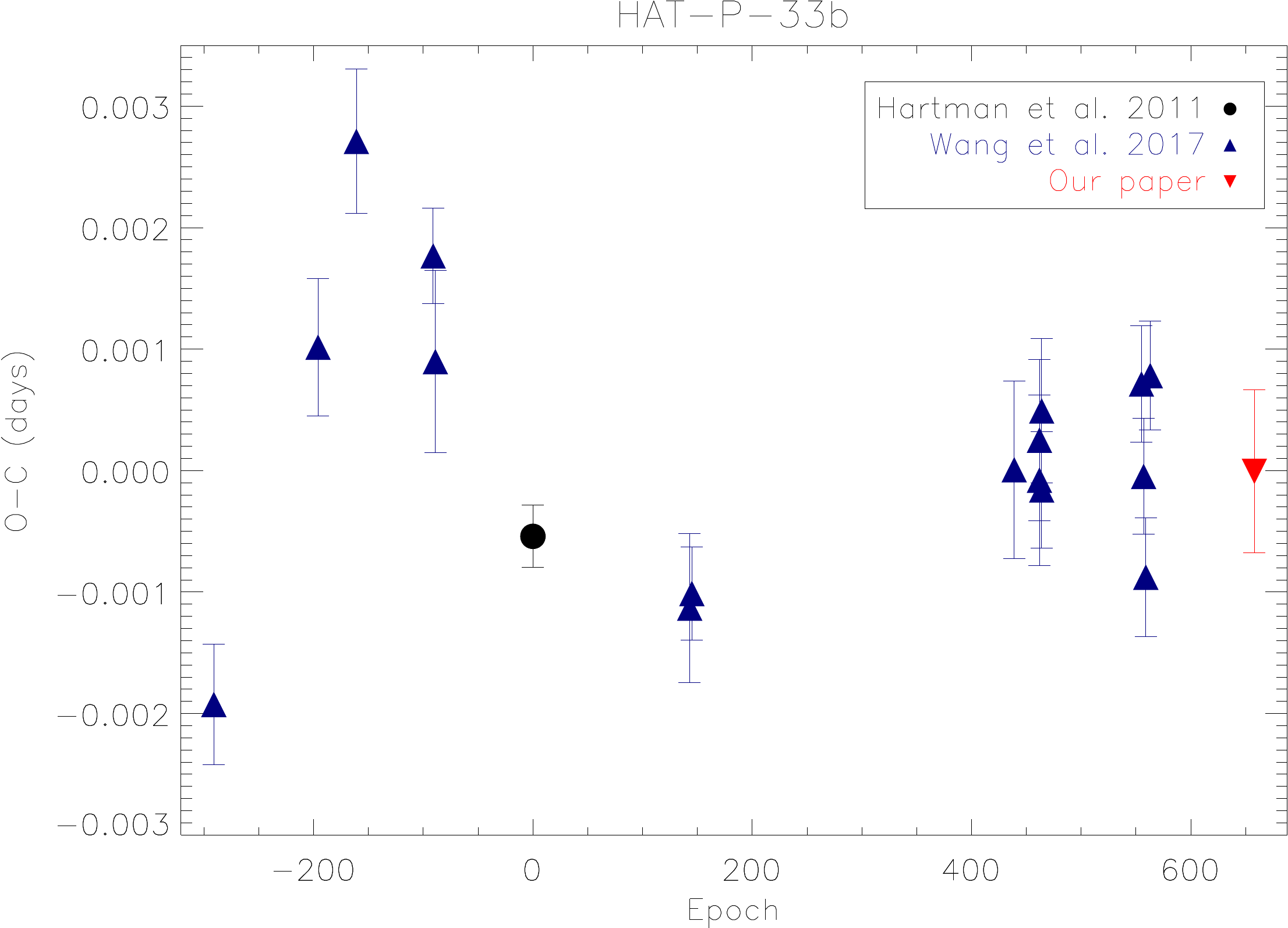} &
\includegraphics[width=0.37\linewidth,angle=90]{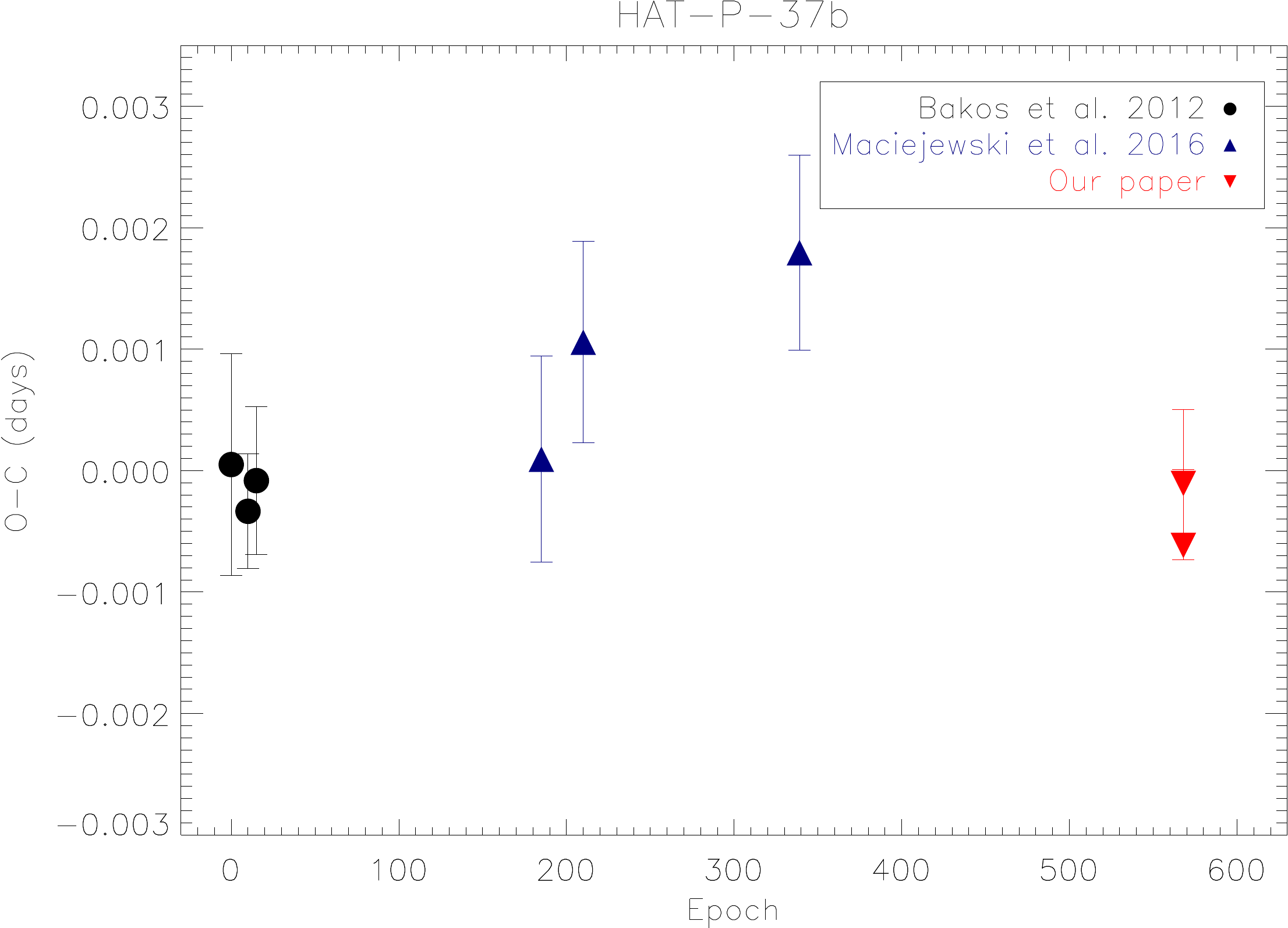}\\
\includegraphics[width=0.37\linewidth,angle=90]{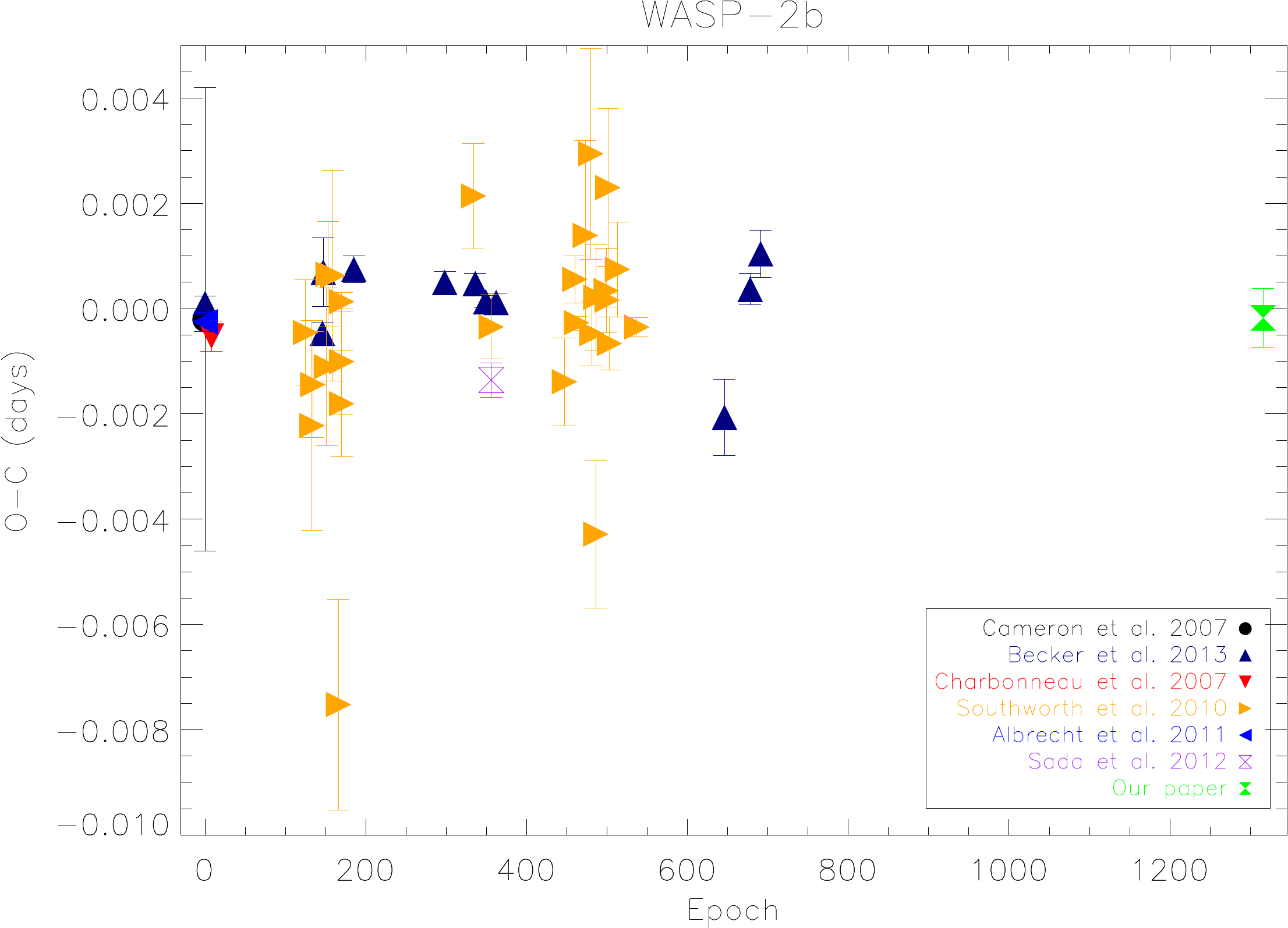} &
\includegraphics[width=0.37\linewidth,angle=90]{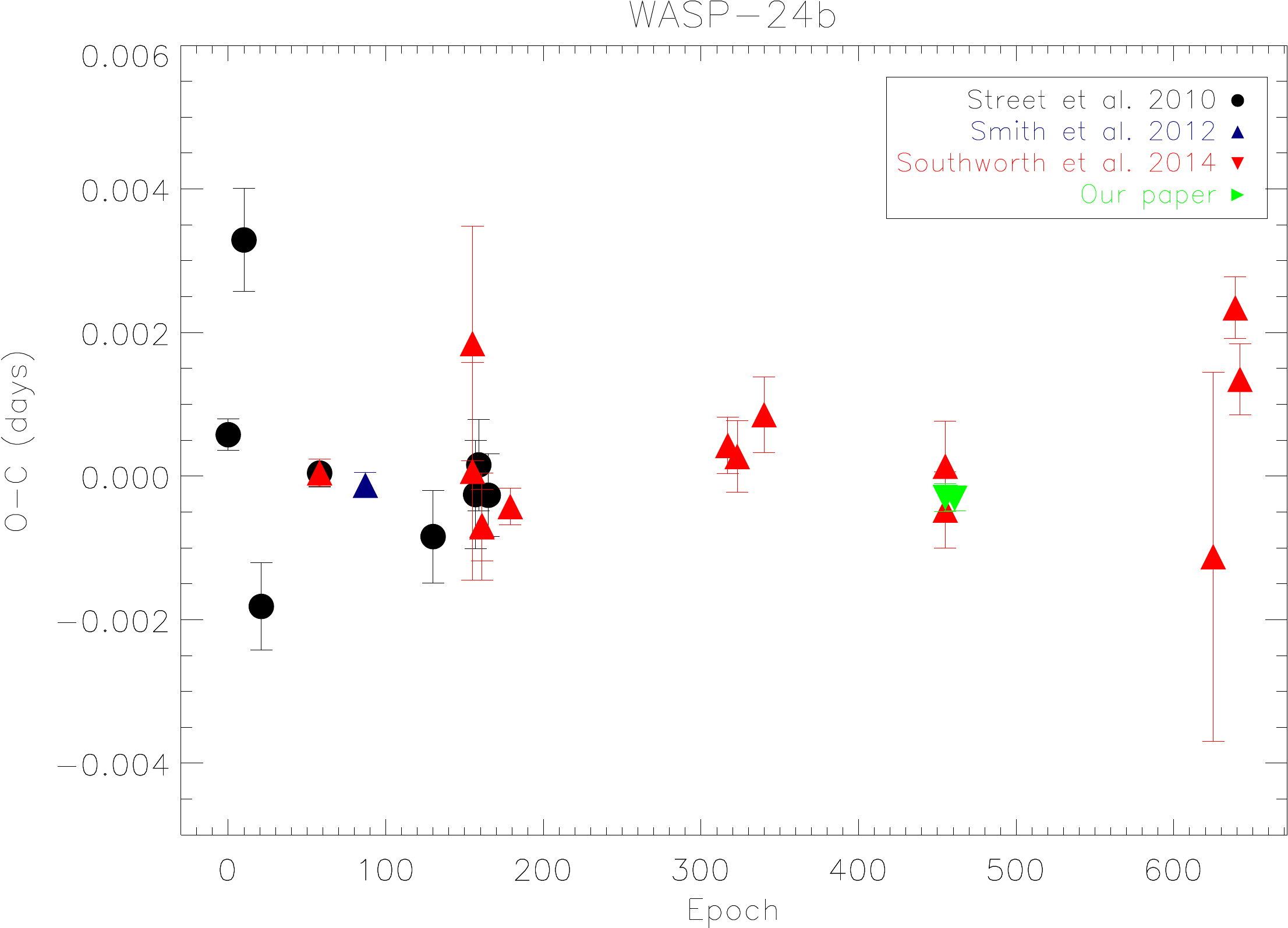} \\
\end{tabular}
\caption{Observation minus calculation mid-transit time (O-C) plots of HAT-P-5b, HAT-P-12b, HAT-P-33b, WASP-2b, and WASP-24b from this paper and previous literature. We do not see any evidence for TTVs.}
\label{fig:oc_1}
\end{figure*}

\begin{figure*}
\centering
\begin{tabular}{c}
\vspace{0.5cm}
\includegraphics[width=0.37\linewidth,angle=90]{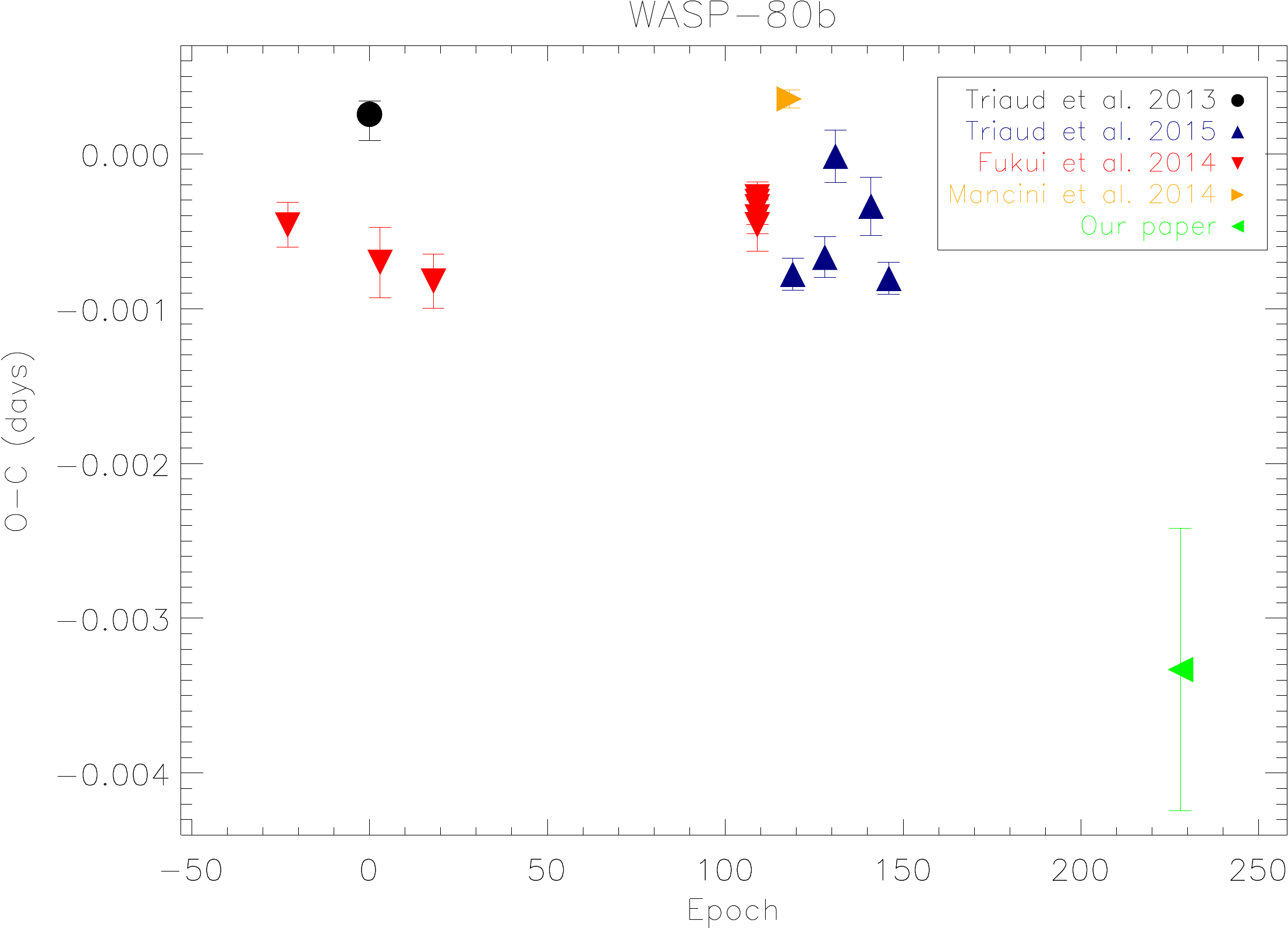} \\
\includegraphics[width=0.37\linewidth,angle=90]{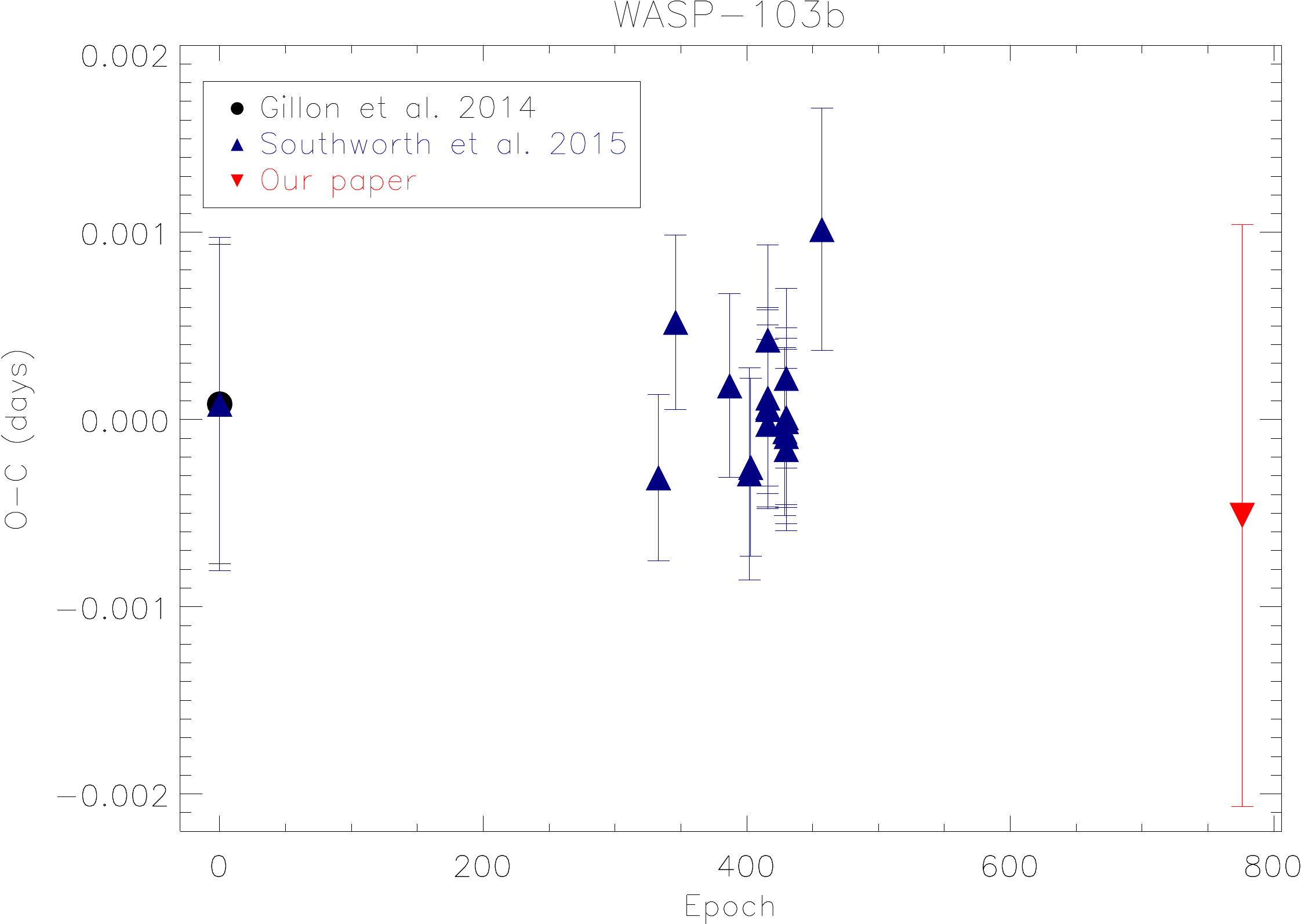}\\
\includegraphics[width=0.37\linewidth,angle=90]{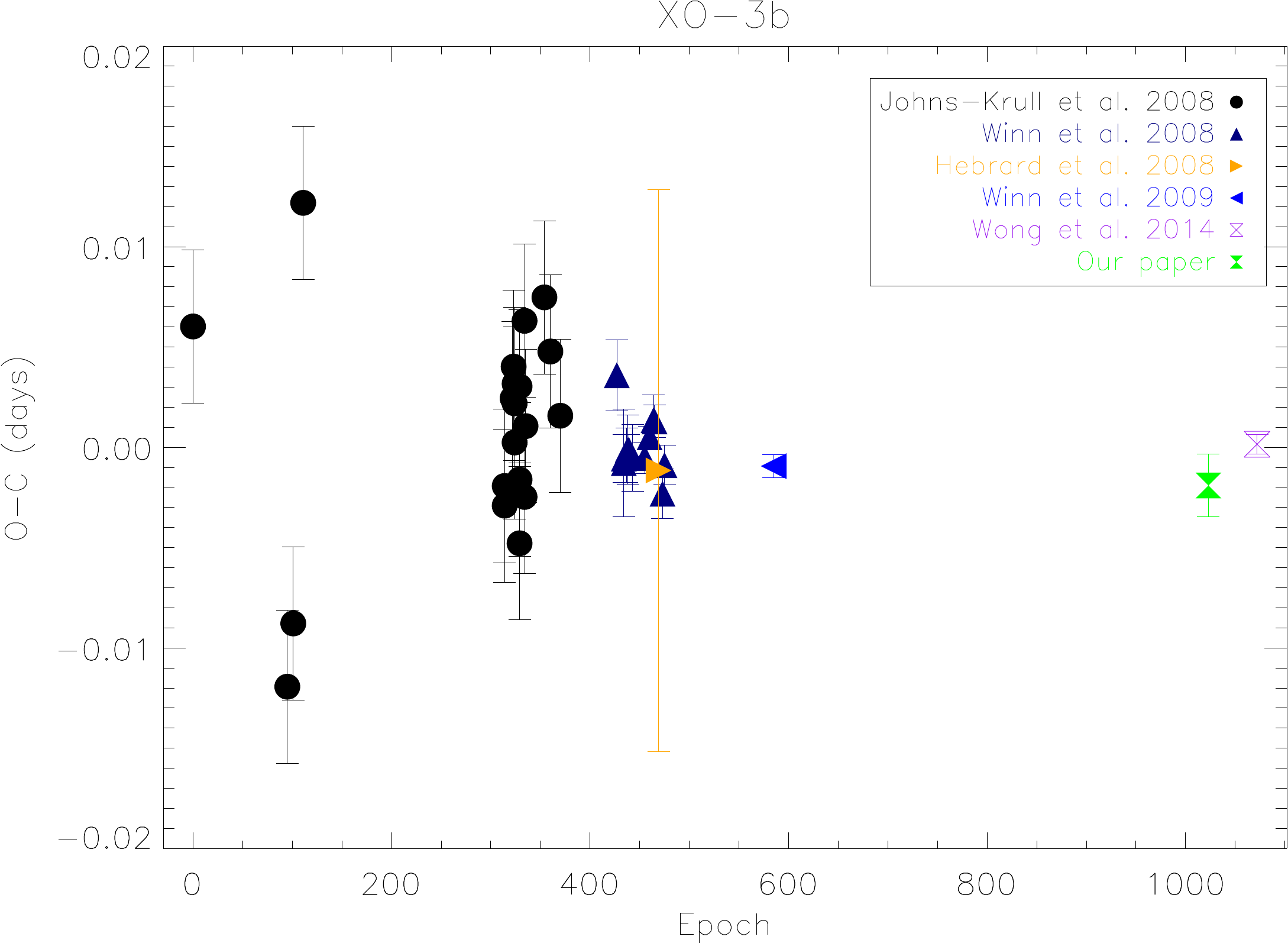}\\
\end{tabular}
\caption{O-C plot of WASP-80b, XO-3b, and WASP-103b from this paper and previous literature. We do not see any TTVs with the exception of WASP-80b but there are large uncertainties in our measurement, so we recommend follow-up observations to verify this claim.}
\label{fig:oc_2}
\end{figure*}

For every system, we also made observation minus calculation mid-transit time (O-C) plots in order to search for any TTVs due to other bodies in the system. We used the derived period and ephemeris found in Tables \ref{tb:pars_1_light}-\ref{tb:pars_2_light} and Equation \ref{eq:Tc} for the calculated mid-transit times. The O-C plots can be found in Figs \ref{fig:oc_1}--\ref{fig:oc_2}. The transit timing analysis for all our targets can be found in Table \ref{tb:mr_timing} (the entire table can be found online). We do not observe any significant TTVs in our data with the exception of a 3.8$\sigma$ deviation for WASP-80b for our observed transit. Since the possible TTV is only one data point and may be caused by an unknown systematic error, more observations of WASP-80b are needed to confirm this result.

\begin{table*}
\centering
\caption{Results of the transit timing analysis$^{1}$}
\begin{tabular}{ccccccc}
\hline
\hline
 Planet Name            	& 
$T_{c}$ ($BJD_{TDB}$) 	&
$T_{c}$ error (d)			&
Epoch				&
O-C (d)				&
O-C error (d)			&
Source 				\\
\hline
\hline
CoRoT-12b				&	 2456338.67097	     & 0.00074 & 686	& -0.0000012		    &0.055 		& This paper \\
CoRoT-12b				&	2454398.628	          &0.055	 & 0		& 0.00000000046		&0.078		& \citealt{Gillon2010}\\
\hline
\end{tabular}
\vspace{-2em}
\tablenotetext{1}{This table is available in its entirety in machine-readable form in the online journal.  A portion is shown here for guidance regarding its form and content. }	
\label{tb:mr_timing}	
\end{table*} 
 
\section{Individual Systems}\label{sec:indiv_systems}

\subsection{CoRoT-12b}
CoRoT-12b was discovered by the CoRoT satellite (\citealt{Carone2012}) and was confirmed by follow-up photometry and radial-velocity measurements (\citealt{Gillon2010}). CoRoT-12b is an inflated hot Jupiter with a low density that is well predicted by standard models (\citealt{Fortney2007}) for irradiated planets (\citealt{Gillon2010}).\\
\indent We observed a transit of CoRoT-12b on 2013 February 15 with the Harris R filter (Fig. \ref{fig:light_1}). We find a $R_{p}/R_{\ast}$ value 4.6$\sigma$ greater than the discovery value. Our derived physical parameters are in good agreement with \citet{Gillon2010}. We find a planetary radius within 1.3$\sigma$ of the previously calculated value and a planetary mass within 1$\sigma$ (Tables \ref{tb:1_light} and \ref{tb:pars_1_light}). 


\subsection{HAT-P-5b}
HAT-P-5b is a hot Jupiter discovered by the HATNet project that orbits a slightly metal-rich star (\citealt{Bakos2007}). Follow-up multi-color transit observations of HAT-P-5b by \citet{Southworth2012b} confirmed the existence of the planet and searched for a variation in planetary radius with wavelength. A significantly larger radius was found in the U-band than expected from Rayleigh scattering alone, which the authors suggest may be due to an unknown systematic error.

\begin{figure*}
\centering
\begin{tabular}{cc}
\vspace{0.5cm}
\includegraphics[width=0.46\linewidth]{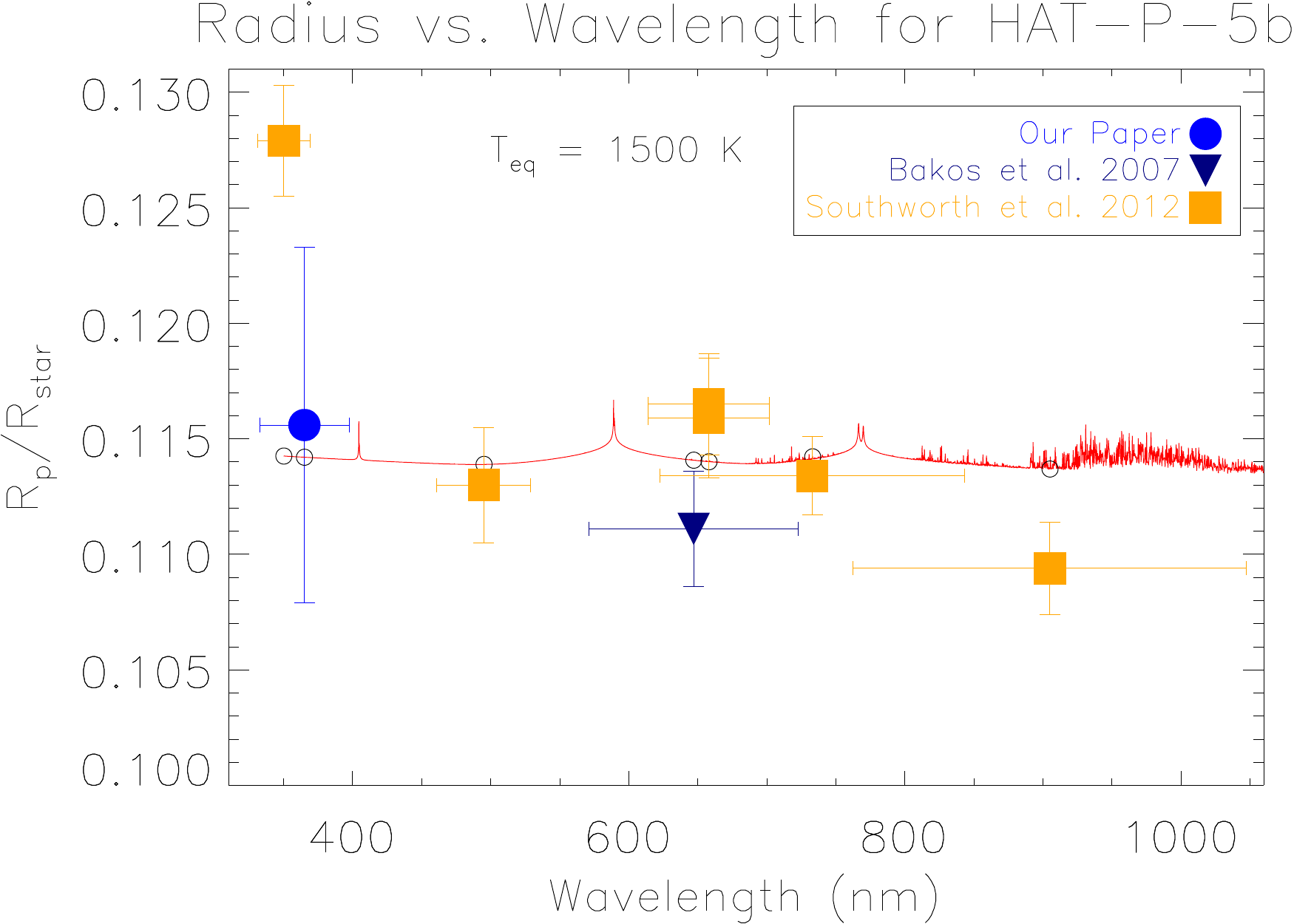} &
\includegraphics[width=0.47\linewidth]{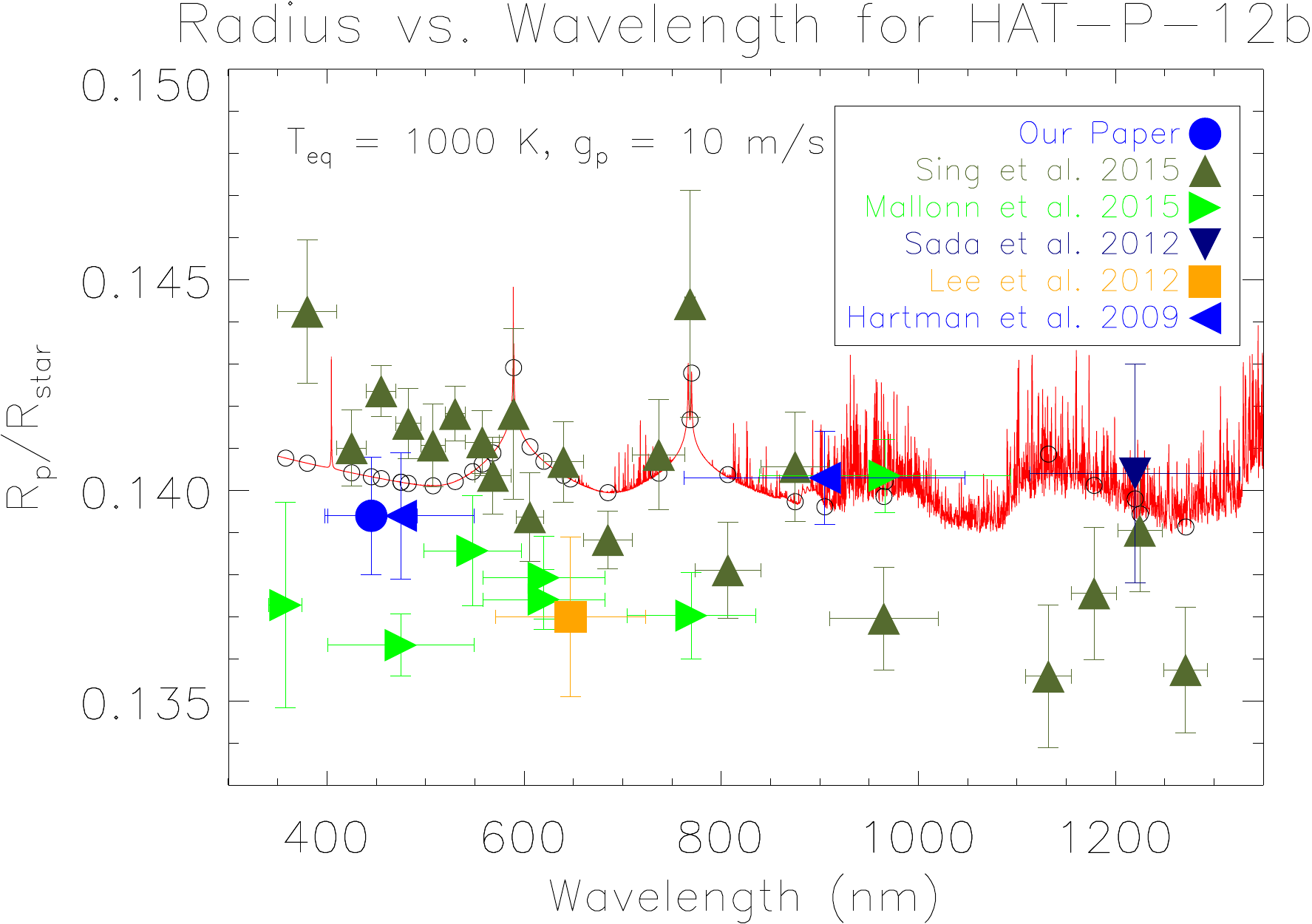} \\
\vspace{0.5cm}
\includegraphics[width=0.46\linewidth]{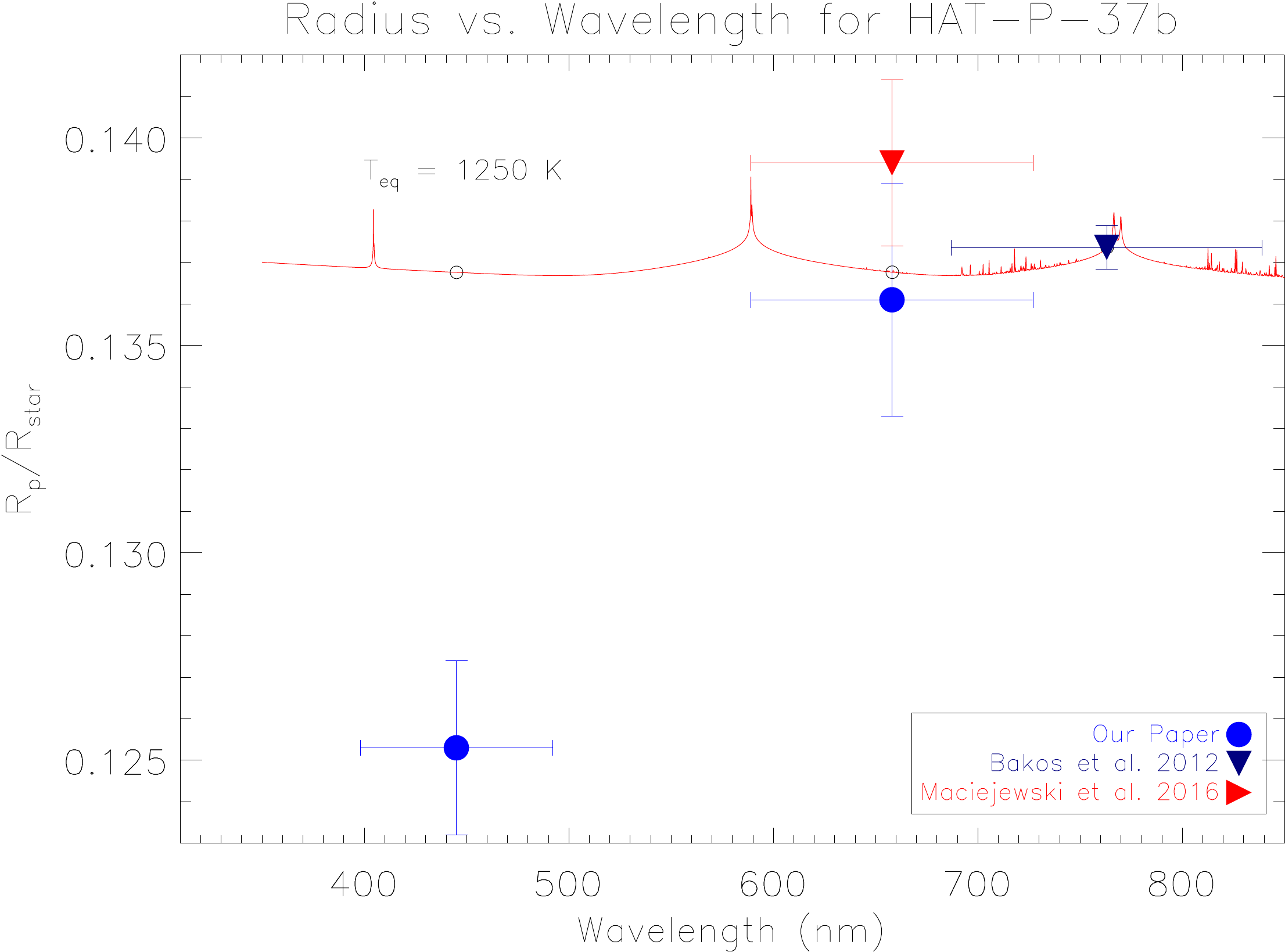} &
\includegraphics[width=0.46\linewidth]{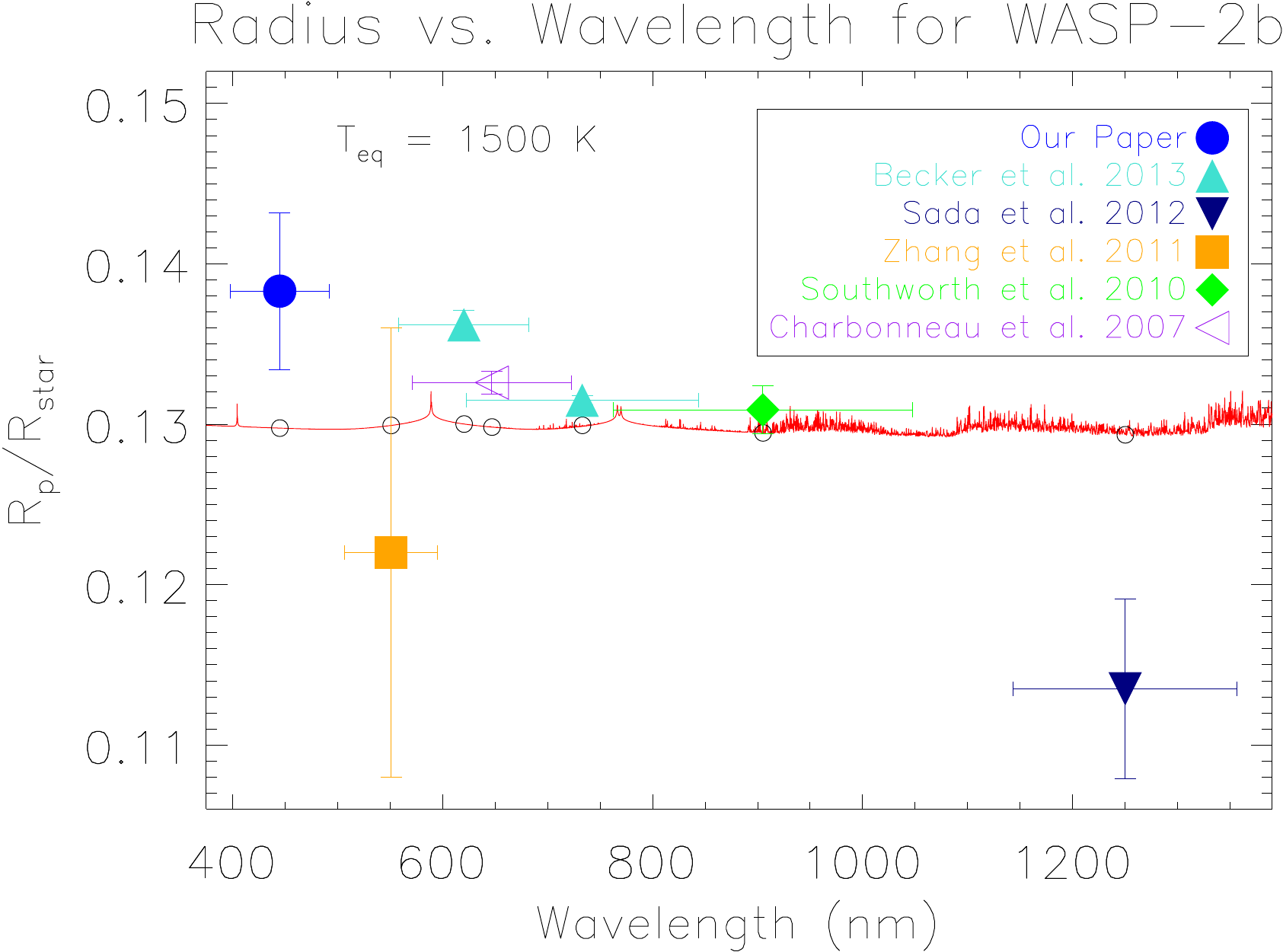} \\
\vspace{0.5cm}
\includegraphics[width=0.47\linewidth]{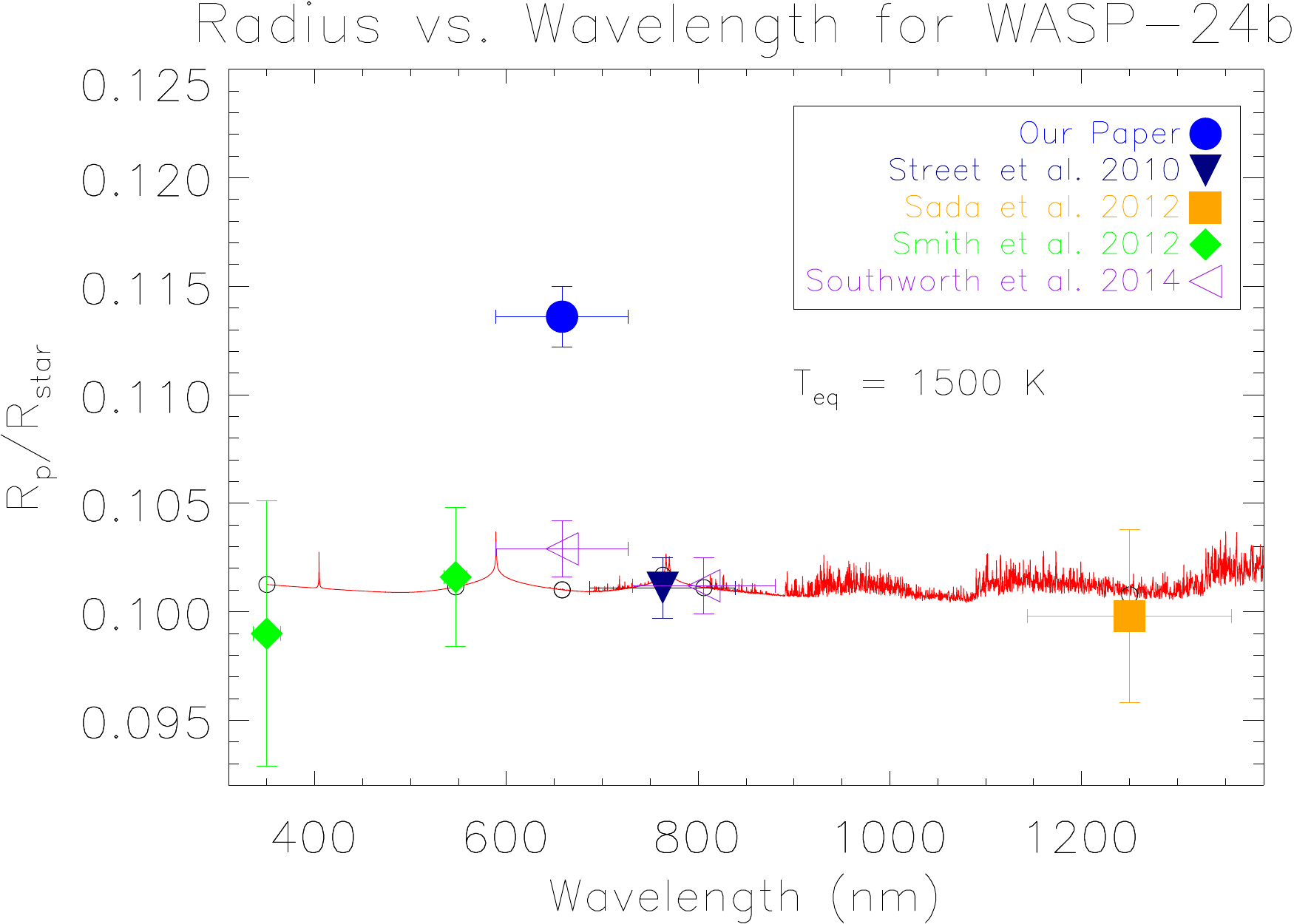} &
\includegraphics[width=0.47\linewidth]{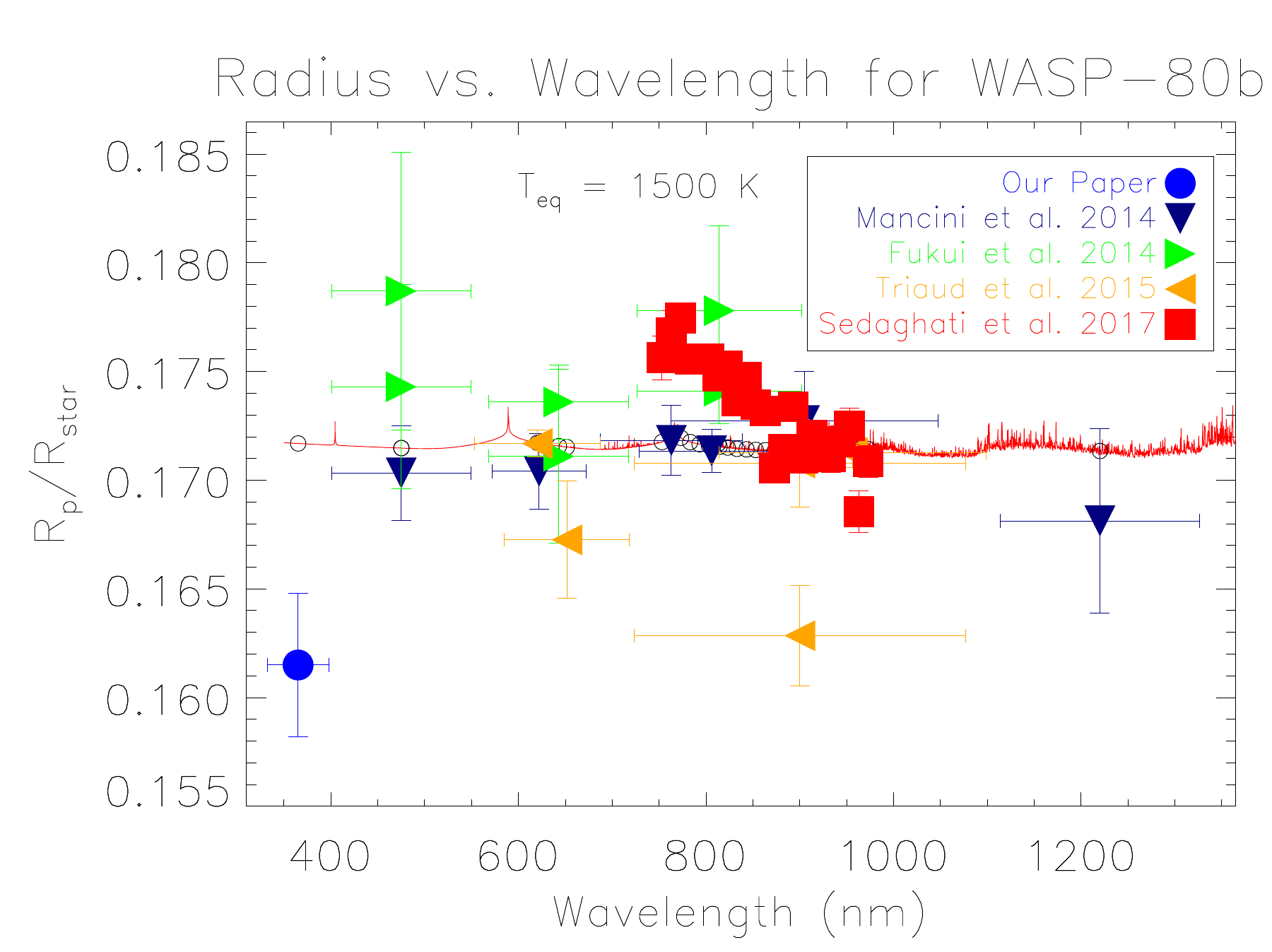} \\
\end{tabular}
\caption{Plot of $R_{p}/R_{*}$ against wavelength for HAT-P-5b, HAT-P-12b, HAT-P-37b, WASP-2b, WASP-24b, and WASP-80b from this paper and previous literature. Our data are shown as blue circles. Over-plotted in red are atmospheric models by \citet{Fortney2010} for planets with a 1 $M_{Jup}$, $g_{p} = 25 m \ s^{-1}$ (unless specified on plot), $T_{eq}$ (specified on plot), and a base radius of 1.25 $R_{Jup}$ at 10 bar. We find that HAT-P-5b, HAT-P-12b, WASP-2b, and WASP-80b have flat spectra that could indicate the presence of clouds. The transit depth variation of HAT-P-37b could be due to absorption of TiO/VO (\citealt{Evans2016}). }
\label{fig:rprstar_1}
\end{figure*}

We observed a transit of HAT-P-5b on 2015 June 6 with the Bessell U filter (Fig. \ref{fig:light_1}). Our derived physical parameters are in agreement with previous literature (Tables \ref{tb:1_light} and \ref{tb:pars_1_light}). We derive a U-band radius consistent with a weighted average of radii taken from 350-733 nm within 1$\sigma$ (Table \ref{tb:rp_rstar}; Fig \ref{fig:rprstar_1}). The error on our U-band observation is too large to determine if the observation by \citet{Southworth2012b} in the same band may have an unknown systematic error (as suggested by them). Our calculated period is in good agreement with the value found by \citet{Southworth2012b} with a similar uncertainty. 

\begin{table*}
\centering
\caption{$R_{p}$/$R_{*}$ and effective central wavelength $\lambda_\textrm{eff}$ from this paper and previous literature for all targets$^{1}$.} 
\begin{tabular}[t]{ccccc}
\hline
Planet & Filter & Wavelength (nm) & Rp/$R_{*}$  & Source \\
\hline
CoRoT-12b & Harris R & 658     & 0.1645$\pm$0.0040 & This Paper \\
CoRoT-12b &Clear    & 400-900 &0.1321$\pm$0.0011   &  \citealt{Gillon2010}\\
\hline
\end{tabular}
\vspace{-2em}
\tablenotetext{1}{This table is available in its entirety in machine-readable form in the online journal.  A portion is shown here for guidance regarding its form and content. }
\label{tb:rp_rstar}
\end{table*}




\subsection{HAT-P-12b}
HAT-P-12b is a low density, sub-Saturn mass planet discovered by the HAT survey (\citealt{Hartman2009}). Multiple photometric studies have further refined the system's parameters and searched for TTVs (\citealt{Sada2012}; \citealt{Sokov2012}; \citealt{Lee2012}; \citealt{Line2013b}; \citealt{Mallonn2015}; \citealt{Sing2016}; \citealt{Sada2016}). \citet{Sing2016} find a strong optical scattering slope from blue to near-IR wavelengths using \textit{Hubble Space Telescope} and \textit{Spitzer Space Telescope} transmission spectrum data. 





We observed a transit of HAT-P-12b on 2014 January 19 using the Harris B filter (Fig. \ref{fig:light_1}). We derive an optical R$_p$/R$_\ast$ within 1$\sigma$ of previously derived radii at optical wavelengths (Table \ref{tb:rp_rstar}; Fig \ref{fig:rprstar_1}). These results are consistent with the planet having high clouds in its atmosphere (e.g. \citealt{Seager2000}; \citealt{Kreidberg2014}) and the finding by \citet{Line2013b} that HAT-P-12b has a cloudy atmosphere. We also find a period similar to \citet{Mallonn2015}. 

\subsection{HAT-P-33b}
HAT-P-33b is an inflated hot Jupiter orbiting a high-jitter star \citep{Hartman2011}. The high-jitter is believed to be caused by convective inhomogeneities in the host star (\citealt{Saar1998}; \citealt{Hartman2011}). The planetary radius and mass, which both depend on eccentricity, and the stellar parameters are not well constrained due to the large jitter ($~$20 \ m \ s$^{-1}$). HAT-P-33b's radius is either 1.7 or 1.8 $R_{Jup}$ assuming a circular or eccentric orbit, respectively. The first follow-up observations by the Transiting Exoplanet Monitoring Project (TEMP) of HAT-P-33b confirmed the discovery parameters and detected no signs of TTVs (\citealt{Wang2017}).

We observed one transit of HAT-P-33b on 2012 April 6 with the Harris R filter (Fig. \ref{fig:light_1}). We find a R-band R$_p$/R$_\ast$ value that is larger by 3.4$\sigma$ from the discovery R$_p$/R$_\ast$ (Table \ref{tb:rp_rstar}). Follow-up observations are need to determine the cause of this discrepancy.


\subsection{HAT-P-37b}
HAT-P-37b was identified by the HATNet survey and was confirmed by high-resolution spectroscopy and further photometric observations (\citealt{Bakos2012}). HAT-P-37b is a hot Jupiter with a planetary mass of 1.169$\pm$0.103 $M_{Jup}$, a radius of 1.178$\pm$0.077 $R_{Jup}$, and a period of 2.797436$\pm$0.000007 $d$. Additional follow-up observations by \citet{Maciejewski2016} confirmed these planetary parameters.

We obtained two transits of HAT-P-37b on 2015 July 1 with the Harris B and R filters (Fig. \ref{fig:light_1}). We derive an R$_p$/R$_\ast$ for each filter that differ by 1.7$\sigma$, with a larger radius in the R band (Table \ref{tb:rp_rstar}; Fig \ref{fig:rprstar_1}). The B-band R$_p$/R$_\ast$ is smaller by 2.85$\sigma$ from the near-IR R$_p$/R$_\ast$ (Table \ref{tb:rp_rstar}; \citealt{Bakos2012}). Our derived R-band R$_p$/R$_\ast$ value agrees within 1$\sigma$ of the Sloan \textit{i} band value obtained by \citet{Bakos2012}. Near-UV observations are needed to determine if the slope between the B and R filters is real or an unknown systematic in the data. Our other derived physical parameters agree with previous literature to within 1$\sigma$ (Tables \ref{tb:1_light} and \ref{tb:pars_1_light}). We also calculate a refined period with a factor of 6 decrease in error. 




\subsection{WASP-2b}
WASP-2b is a short-period hot Jupiter discovered by the WASP survey and confirmed by radial-velocity measurements taken with the SOPHIE spectrograph (\citealt{Cameron2007}). Extensive photometry and radial velocity measurements have been performed on WASP-2b, further refining its system parameters (\citealt{Charbonneau2007}; \citealt{Daemgen2009}; \citealt{Southworth2010b}; \citealt{Triaud2010}; \citealt{Albrecht2011}; \citealt{Zhang2011}; \citealt{Husnoo2012}; \citealt{Sada2012}; \citealt{Becker2013}).

We observed WASP-2b on 2014 June 14 with the Harris B filter (Fig. \ref{fig:light_2}). Our derived physical parameters and transit depth agree with previous literature to within 1$\sigma$ and we calculate a period with a factor of 2 decrease in error (Tables \ref{tb:1_light} and \ref{tb:pars_1_light}). 

\subsection{WASP-24b}
WASP-24b is a hot Jupiter detected by WASP and confirmed by radial velocity measurements and additional photometric observations (\citealt{Street2010}). Further photometric studies calculated improved system parameters (\citealt{Southworth2014}) and radial velocity measurements were used to determine that the planet exhibits a symmetrical Rossiter-McLaughlin effect, indicating a prograde, well-aligned orbit (\citealt{Simpson2011}).

Our observations of WASP-24b took place on 2012 March 23 and 2012 April 6 (Fig. \ref{fig:light_2}). We obtained two transits with the Harris R filter and each transit was modeled separately. The R$_p$/R$_\ast$ of both dates overlap each other within 1$\sigma$. We then found the weighted average of the light curve parameters before deriving the physical parameters. Our weighted average R$_p$/R$_\ast$ disagrees with previous R-band observations (\citealt{Southworth2014}) by 4$\sigma$ (Table \ref{tb:rp_rstar}; Fig \ref{fig:rprstar_1}). The cause of this difference is unknown but future observations can put better constraints on transit depth and solve this discrepancy. Our other derived parameters generally agree with previous results except for our planetary radius and equilibrium temperature, which differ from \citet{Southworth2014} by 1.4$\sigma$ and 1.9$\sigma$, respectively (Tables \ref{tb:1_light} and \ref{tb:pars_1_light}). We calculate a new period with a factor of 2.6 decrease in error (Table \ref{tb:pars_1_light}).

\subsection{WASP-60b}
WASP-60b was identified by WASP-North and was confirmed by radial-velocity measurements and follow-up photometry (\citealt{Hebrard2013}). WASP-60b is an unexpectedly compact planet orbiting a metal-poor star.

We observed a transit of WASP-60b with the Harris B filter on 2012 December 1 (Fig. \ref{fig:light_2}). This observation is the first follow-up light curve of WASP-60b. During observations, the automatic guider briefly failed, resulting in a hole in the transit light curve. Despite this, we are able to derive parameters that agree with previous literature to within 1$\sigma$ (Tables \ref{tb:1_light} and \ref{tb:pars_1_light}). We find a B-band $R_{p}/R_{\ast}$ value 1.3$\sigma$ greater than the discovery $R_{p}/R_{\ast}$.  



\subsection{WASP-80b}
WASP-80b is a warm Saturn/hot Jupiter ($M_{p}= 0.55\pm0.04M_{jup}$) with one of the largest transit depths (0.17126$\pm$0.00031) discovered so far (\citealt{Triaud2013}). Multiple photometric studies have been done at various wavelengths to refine WASP-80b's planetary parameters (\citealt{Fukui2014}; \citealt{Mancini2014}; \citealt{Triaud2015}; \citealt{Salz2015}; \citealt{Sedaghati2017}). The planet has a transmission spectrum consistent with thick clouds and atmospheric haze (\citealt{Fukui2014}).

We observed WASP-80b on 2014 June 16 with the Bessell U filter, obtaining one transit (Fig. \ref{fig:light_2}). Inclement weather conditions caused our guider to briefly fail, resulting in a hole in the transit light curve. We derive physical parameters that closely agree with previous literature and also calculate a slightly a refined period with a factor of 2 decrease in error (Tables \ref{tb:1_light} and \ref{tb:pars_1_light}). Our observations possibly detect a TTV compared to previous work by 3.7$\sigma$ (Section \ref{sec:period}), however, further observations of WASP-80b are needed in order to confirm this result.

\subsection{WASP-103b} \label{sec:wasp103b}
WASP-103b is a hot Jupiter detected by the WASP survey with a mass of 1.49$\pm$0.09 $M_{jup}$, short period planet ($P_{p} = 0.925542\pm0.000019 \ \rm{d}$), and has an orbital radius only 20\% larger than its Roche radius (\citealt{Gillon2014}). It was found that there is a faint, cool, and nearby (with a sky-projected separation of 0.242$\pm$0.016 arcsec) companion star of WASP-103 (\citealt{Wollert2015}; \citealt{Ngo2016}). Further photometric observations were made to refine WASP-103b's planetary parameters and ephemeris (\citealt{Southworth2015}; \citealt{Southworth2016}; \citealt{Lendl2017}). A comparison of observed planetary radius at different wavelengths found a larger radius at bluer optical wavelengths, but \citet{Southworth2016} state that Rayleigh scattering cannot be the main cause even when including the contamination of the nearby companion star. 

\begin{figure*}
\centering
\begin{tabular}{cc}
\vspace{0.5cm}
\includegraphics[width=0.47\linewidth]{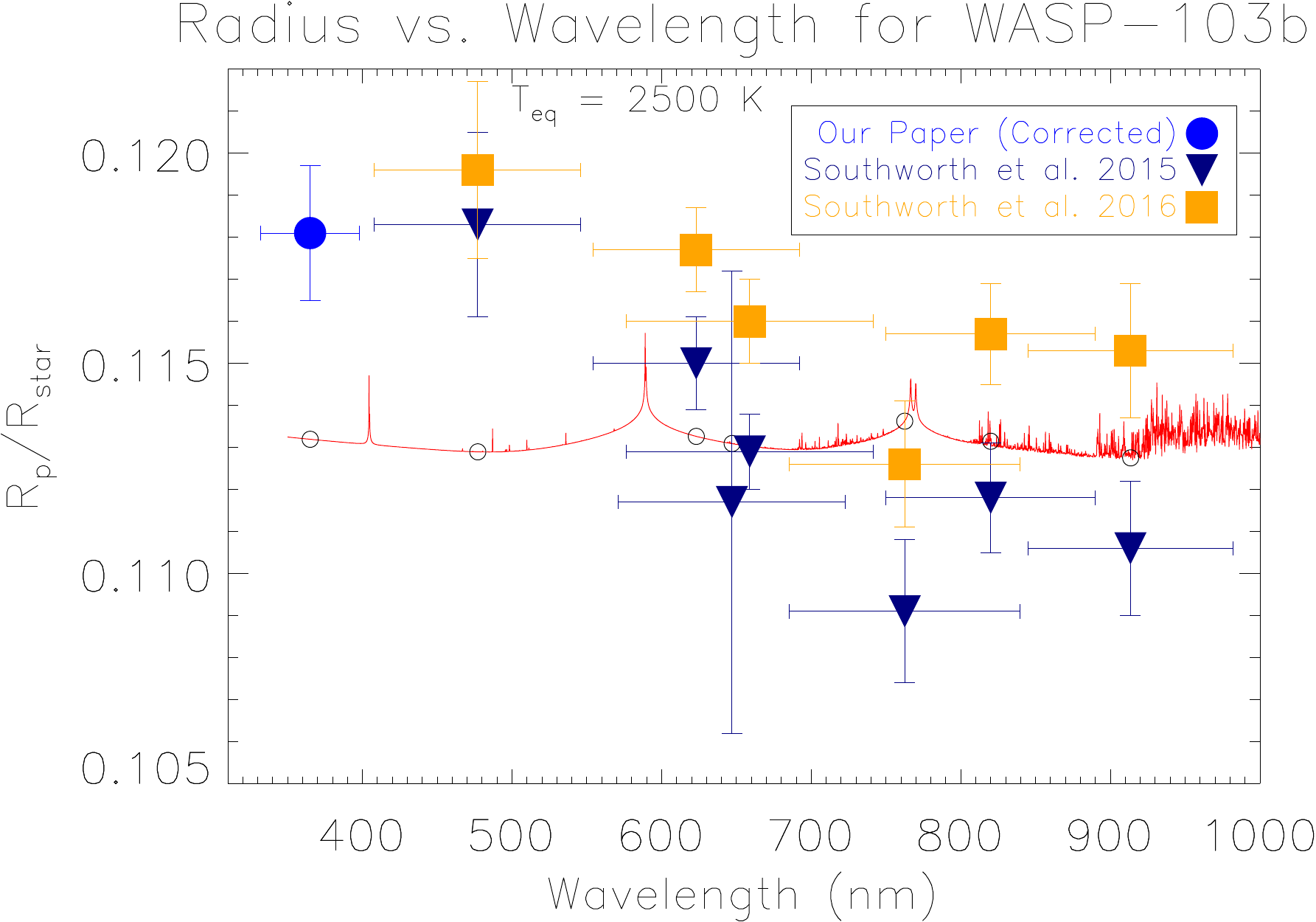} &
\includegraphics[width=0.46\linewidth]{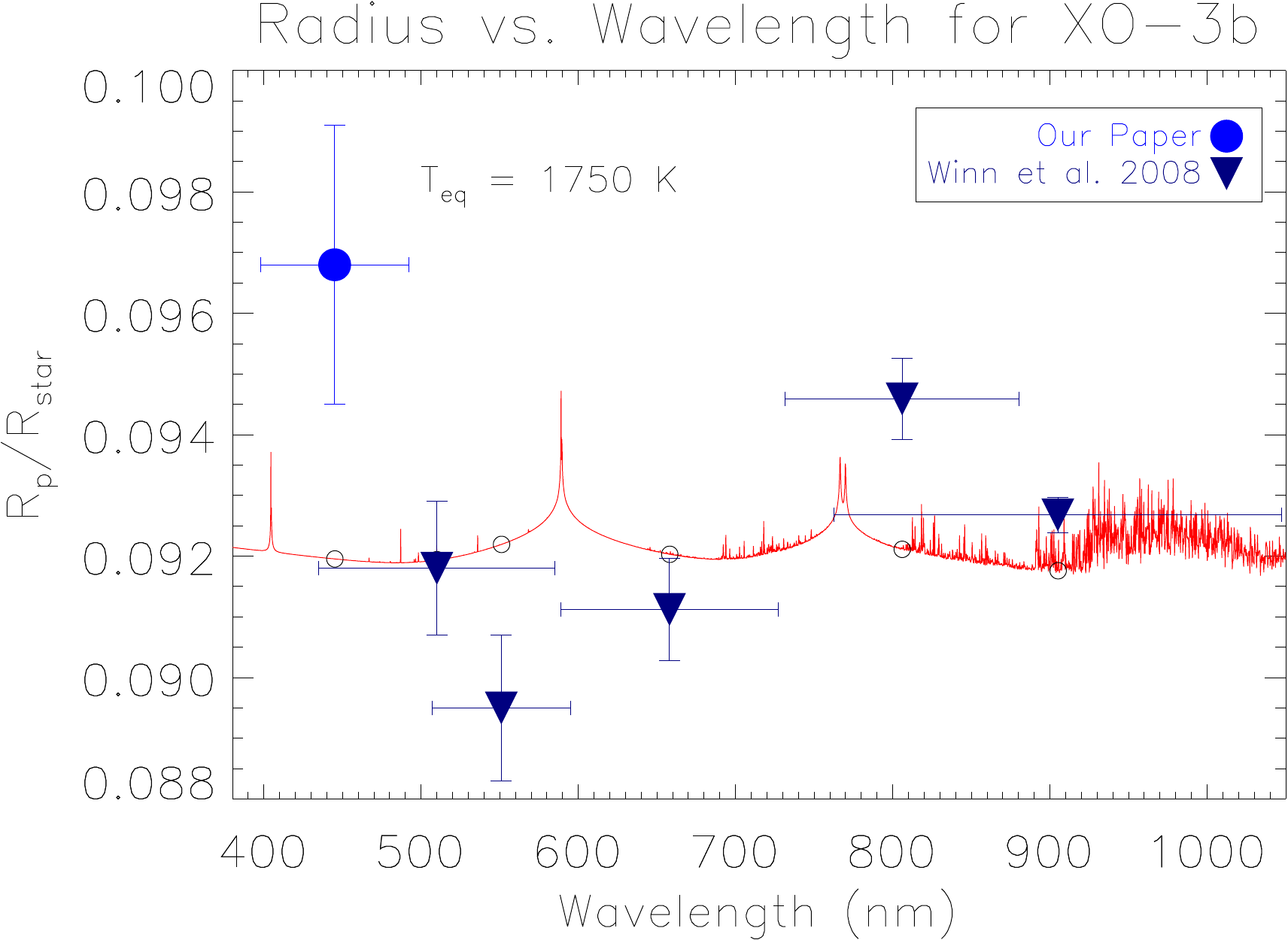} \\
\end{tabular}
\caption{Plot of Rp/$R_{*}$ against wavelength for WASP-103b and XO-3b from this paper and previous literature. Both  WASP-103b and XO-3b show variations with wavelength. Other comments are the same as Fig. \ref{fig:rprstar_1} }
\label{fig:rprstar_2}
\end{figure*}

We observed WASP-103b on 2015 June 3 with the Bessell U filter (Fig. \ref{fig:light_2}). We derive a dilution-corrected near-UV $\left( R_{p}/R_{\ast} \right)_{cor}$ that differs from the discovery value by 2.1$\sigma$. Our other calculated parameters agree with previous literature to within 1$\sigma$ and our calculated period closely agrees with the period found by \citet{Southworth2015} (Tables \ref{tb:1_light} and \ref{tb:pars_2_light}). A variation in R$_p$/R$_\ast$ is found from the ultraviolet to the near-infrared wavelengths (Table \ref{tb:rp_rstar}; Fig \ref{fig:rprstar_2}) consistent with that found by \citet{Southworth2016}. 

We correct for the dilution due to the companion star being in our aperture using the procedure described below (this procedure is similar to that done by \citealt{Southworth2016}). (1) The light curve is modeled with EXOMOP and we find an uncorrected transit depth of $\left(R_p/R_\ast\right)_{uncor}$ = 0.1174$\pm$0.0016. (2) Theoretical spectra of both stars is produced using ATLAS9-ODFNEW (\citealt{Castelli2004}). For WASP-103 we use $T_\textrm{eff}$ = 6110 K and $M_{star} = 1.22 \ M_{\sun}$ (\citealt{Gillon2014}) and for the companion star we use $T_\textrm{eff}$ = 4405 K (\citealt{Southworth2016}) and $M_{star} = 0.721 \ M_{\sun}$ (\citealt{Ngo2016}). Additionally, in order to scale the spectrum correctly we use the mass-lumnosity relation $L=L_{\sun}\left(M/M_{\sun}\right)^4$ for stars between 0.5 and 2 $M_{\sun}$. (3) The ATLAS9-ODFNEW model spectra is convolved with the bandpass of the Bessell U filter (\citealt{Bessell1979}). (4) The corrected transit depth, $\left( R_{p}/R_{\ast} \right)_{cor}$, is found using the equation (\citealt{Ciardi2015})
\begin{equation}
\left(\frac{R_{p}}{R_{\ast}}\right)_{cor}    = \left(\frac{R_p}{R_\ast}\right)_{uncor} \sqrt{\frac{F_{tot}}{F_{2}} }, 
\end{equation}
where $F_{tot}$ is the total flux of both stars and $F_{2}$ is the flux from the companion star. In \citet{Southworth2016} the error of the photometric light curve dominated the error calculation of their corrected transit depth and therefore we also use our photometric error bars for the error in the $(R_{p}/R_{\ast})_{cor}$. Using this procedure we find a $(R_{p}/R_{\ast})_{cor}~=~0.1181\pm0.0016$. 

\subsection{XO-3b}
XO-3b is a massive planet (11.79$\pm$0.59 $M_{Jup}$) with a large eccentricity (0.26$\pm$ 0.017) detected by the XO survey (\citealt{Johns-Krull2008}). Further photometric observations have refined the system's parameters (\citealt{Winn2008}; \citealt{Hirano2011b}; \citealt{Machalek2010}; \citealt{Wong2014}) and \citet{Hebrard2008} found that XO-3's spin axis is misaligned with XO-3b's rotation axis.


We observed a transit of XO-3b on 2012 November 30 with the Harris B filter (Fig. \ref{fig:light_3}). We derive physical parameters that are in agreement with previous literature (Tables \ref{tb:1_light} and \ref{tb:pars_2_light}). Our calculated R$_p$/R$_\ast$ is 2$\sigma$ larger than the V-band R$_p$/R$_\ast$ found by \citet{Winn2008}. We calculate a refined period with an error decreased by a factor of 13 from the value found by \citet{Winn2008}. A non-flat spectrum for R$_p$/R$_\ast$ is found for XO-3b (Table \ref{tb:rp_rstar}; Fig \ref{fig:rprstar_2}).

\section{Discussion}\label{sec:discuss}
\subsection{Wavelength dependence on the transit depth} 

We find a constant transit depth across optical wavelengths for the TEPs HAT-P-5b, HAT-P-12b, WASP-2b, WASP-24b, and WASP-80b (Fig \ref{fig:rprstar_1}, Table \ref{tb:rp_rstar}). A lack of variation in radius with wavelength could suggest these planets (HAT-P-5b, HAT-P-12b, WASP-2b,  WASP-80b) have clouds/hazes in their upper atmospheres (e.g. \citealt{Seager2000}; \citealt{Brown2001}; \citealt{Gibson2013b}; \citealt{Marley2013}; \citealt{Kreidberg2014}) or they have an isothermal pressure-temperature profile \citep{Fortney2006}. \citet{Mancini2014} also do not detect a significant variation in WASP-80b's transit depth with wavelength, and \citet{Southworth2012b} finds a relatively flat spectrum of planetary radii for HAT-P-5b with the exception of their observed radius in the U-band (which they suspect is caused by systematic error in their U-band photometry). A flat spectrum for WASP-24b is also found with the exception of one value. Our R-band R$_p$/R$_\ast$ found for WASP-24b differs by 4$\sigma$ from the previously calculated R$_p$/R$_\ast$ (\citealt{Southworth2014}) for that same band. The cause of this is unclear and future observations are needed to investigate. Our results are consistent with other transiting exoplanet observations having a flat spectrum in optical wavelengths (i.e. TrES-3b, \citealt{Turner2013a};  GJ 1214b, \citealt{Bean2011}; \citealt{Kreidberg2014}; WASP-29b, \citealt{Gibson2013a}; \citealt{Gibson2013b}; HAT-P-19b, \citealt{Mallonn2015b}; HAT-P-1b, HAT-P-13b, HAT-P-16b, HAT-P-22b, TrES-2b, WASP-33b, WASP-44b, WASP-48b, WASP-77Ab, \citealt{Turner2016}). 

We find variations in the transit depth with wavelength for CoRoT-12b, HAT-P-33b, HAT-P-37b, WASP-103b, and XO-3b (Fig \ref{fig:rprstar_1}-\ref{fig:rprstar_2}, Table \ref{tb:rp_rstar}), which could indicate scattering (i.e. due to aerosols or Rayleigh scattering) or absorption in their atmospheres (e.g. \citealt{Benneke2012}; \citealt{Griffith2014}). Our observation of HAT-P-37b exhibits a smaller transit depth in B-band than the red/near-IR value. Such a variation has only been seen in a recent paper by \citet{Evans2016} where they observe a smaller B-band transit depth than optical in WASP-121b. \citet{Evans2016} believe a possible cause of such a variation is TiO/VO absorption and this may also be the cause of the transit depth variations seen in HAT-P-37b. However, more theoretic modeling is needed to confirm that TiO/VO is in fact the opacity source. Additionally, a smaller near-UV radius was recently observed in the hot Jupiter WASP-1b (\citealt{Turner2016}), however, these observations did not observe in the B-band. Future near-UV and blue-band observations are needed for WASP-103b and XO-3b to determine whether the scattering in their atmospheres is due to Rayleigh scattering (\citealt{Etangs2008}; \citealt{Tinetti2010}; \citealt{deWit2013}; \citealt{Griffith2014}) since these bands are the only optical wavelengths not affected by strong spectral features. The radius variations in WASP-103b show a consistently larger transit depth in the near-UV and blue than the rest of the optical (this varation is still present when corrected for dilution due the companion star). Such a radius variation may indicate a change in particle size at different altitudes of the planetary atmosphere (e.g. \citealt{Wakeford2015}). We find a larger R-band transit depth in HAT-P-33b and CoRoT-12b than their discovery transit depths. Since the R-filter encompasses the $H\alpha$ line (656.281 nm), our observation could be an indication of atmospheric escape such as that observed in the atmospheres of HD 189733b (\citealt{Jensen2012}; \citealt{Cauley2015}; \citealt{Barnes2016}; \citealt{Cauley2017a}; \citealt{Cauley2017b}) and HD 209458b \citep{Defru2013} and predicted (e.g. \citealt{Christie2013}; \citealt{Turner2016cloudy}). Follow-up photometry and high-resolution spectroscopy observations are encouraged to confirm all the transit depth variations. These results also agree with observations of other exoplanets not having a flat spectrum (i.e. HD 209458b, \citealt{Sing2008}; HAT-P-5b, \citealt{Southworth2012b}; GJ 3470b, \citealt{Nascimbeni2013}; Qatar-2, \citealt{Mancini2014}; WASP-17b, WASP-39b, HAT-P-1b, WASP-31b, HAT-P-12b, HD189733b, WASP-6b, \citealt{Sing2016}; CoRoT-1b, TrES-4b, WASP-1b, WASP-12b, WASP-36b, \citealt{Turner2016}).


For illustration, the observed R$_p$/R$_\ast$ differences with wavelength for each target (Table \ref{tb:rp_rstar}) are compared to theoretical predictions (\citealt{Fortney2010}) for a model planetary atmosphere (Figure \ref{fig:rprstar_1}--\ref{fig:rprstar_2}). The models used are calculated for planets with a 1 $M_{Jup}$, $g_{p} = 25 m s^{-1}$ or $g_{p} = 10 m s^{-1}$, base radius of 1.25 $R_{Jup}$ at 10 bar, $T_{eq}$ closest to the measured value for each exoplanet (with model choices of 500, 750, 1000, 1250, 1500, 1750, 2000, 2500 K), and solar metallicity. To provide a best fit to the spectral changes a vertical offset is applied to the model. This comparison is helpful as it illustrates the size of observed variation compared to what the theoretical models predict. However, radiative transfer models calculated for each exoplanet individually are needed to fully understand their transmission spectra.

 Finally, no signs of asymmetric transits are seen in the near-UV light curves of HAT-P-5b, WASP-80b, and WASP-103b. This result is consistent with ground-based near-UV observations of 19 other transiting exoplanets (\citealt{Southworth2012b}; \citealt{Pearson2014}; \citealt{Turner2013a}; \citealt{Bento2014}; \citealt{Copperwheat2013}; \citealt{Zellem2015}; \citealt{Turner2016}) that show no evidence of asymmetric transits. Additionally, theoretical modeling by \citet{Turner2016cloudy} using the \texttt{CLOUDY} plasma simulation code showed that asymmetric transits cannot be produced in the broad-band near-UV band regardless of the assumed physical phenomena that could cause absorption (e.g. \citealt{Vidotto2010}; \citealt{Lai2010}; \citealt{BenJaffel2014}; \citealt{Matsakos2015}; \citealt{Kislyakova2016}).   

\subsubsection{Variability in the host stars} \label{sec:var}



One of the major assumptions in our interpretation that the planetary atmosphere is the cause of the transit depth variations is that the brightness of the host stars have minimal variability due to stellar activity. The presence of star spots and stellar activity can produce variations in the observed transit depth (e.g. \citealt{Czesla2009}; \citealt{Oshagh2013}; \citealt{Oshagh2014}; \citealt{Zellem2015}; \citealt{Zellem2017}). This effect is stronger in the near-UV and blue and can mimic a Rayleigh scattering signature (e.g. \citealt{Oshagh2014}; \citealt{McCullough2014}). Additionally, no obvious star spot crossing is seen in our data (Figs. \ref{fig:light_1}-\ref{fig:light_3}) with the possible exception of HAT-P-37b (see below).

We estimate how much the transit depth may change due to unocculted spots using the formalization presented by \citet{Sing2011}. This method assumes that the spots can be treated as a stellar spectrum but with a lower effective temperature, no surface brightness variation outside the spots, and no plage are present. The effect of these assumptions are a dimming of the star and therefore an increase in the transit depth. \citet{Sing2011} find for HD 189733b that the change in transit depth due to unocculted spots, $\Delta \left(R_{p}/R_\ast\right) = 2.08\times10^{-3}/2 \left(R_{p}/R_\ast\right)$ between 375--400 $\rm{nm}$. Therefore, unocculted spots have minimal influence (assuming our host stars have unocculted spots similar to HD 189733b) on the observed transit depth variations since our final error bars (Table \ref{tb:1_light}) are at least 10 times larger than the influence of these spots (e.g. $\Delta \left[R_{p}/R_\ast\right]$ = 0.00014 for HAT-P-37b). Qualitatively, this result is consistent with the study by \citet{Llama2015} that find that stellar activity similar to the sun has very little effect on the transit depth measured in near-UV to optical wavelengths. Nonetheless, we highly encourage follow-up observations and host star monitoring of all our targets to assess the effect of stellar activity on the observed transit depth variations. 

Next, we investigate what effect a star-spot crossing in the light curve of HAT-P-37 would have on its calculated transit depth. In the B-band light curve of HAT-P-37b (Figure \ref{fig:light_1}) there may be a star-spot crossing at a phase range of 0.004--0.008. However, the detected signal is very close to the scatter in the light curve. If we model the light curve without the possible star-spot crossing we find a $\left(R_{p}/R_\ast\right) = 0.1278\pm0.0048$, within 1$\sigma$ of the transit depth of the entire light curve (0.1253$\pm0.0021$). \citet{McCullough2014} present a procedure to estimate the effects of unocculted spots on the transit depth. Their procedure can also be used to estimate the effect of star spot crossings on the transit depth, where instead of unocculted spots increasing the transit depth occulted spots should decrease it. \citet{McCullough2014} find that the change in transit depth due to spots, $\Delta(R_{p}^2/R_\ast^2)$, is
\begin{equation}
\Delta \left(\frac{R_{p}^2}{R_{\ast}^2}\right) = \left( \frac{R_{p}}{R_\ast} \right)^2 \delta \frac{T_{spot}}{T_\textrm{eff}},
\end{equation}
where $R_{p}/R_{\ast}$ is the unperturbed transit depth, $\delta$ is the fractional area of star spots, and $T_{spot}$ is the temperature of the spot. If we set $\Delta(R_{p}^2/R_\ast^2)$ = 3100 ppm (the approximate difference between our B-band and the Sloan-i transit depth; \citealt{Hartman2011}), then we can estimate $T_{spot}$ and $\delta$. For spot temperatures between 2000--5000 K, we find that $\delta$ would be between 20 - 50 $\%$. Typical values of $\delta$ for solar-like stars is around several $\%$ (e.g. \citealt{Pont2008}; \citealt{Sing2011}; \citealt{McCullough2014}), so our estimated $\delta$ range is extremely high. Due to both these tests, it seems unlikely that the smaller B-band transit depth of HAT-P-37b is due to an occulted star-spot.

\section{Conclusions}
We observed 11 transiting hot Jupiters (CoRoT-12b, HAT-P-5b, HAT-P-12b, HAT-P-33b, HAT-P-37b, WASP-2b, WASP-24b, WASP-60b, WASP-80b, WASP-103b, XO-3b) from the ground using near-UV and optical filters in order to update their system parameters and constrain their atmospheres. Our observations of CoRoT-12b, HAT-P-37b and WASP-60b are the first follow-up observations of these planets since their discovery and we also obtain the first near-UV light curves of WASP-80b and WASP-103b. We find that HAT-P-5b, HAT-P-12b, WASP-2b, WASP-24b, and WASP-80b exhibit a flat spectrum across the optical wavelengths, suggestive of clouds in their atmospheres. Variation in the transit depths is observed for WASP-103b and XO-3b and may indicate scattering in their atmospheres. Additionally, we observe a smaller B-band transit depth compared to near-IR in HAT-P-37b. Such a variation may be caused by TiO/VO absorption (\citealt{Evans2016}). We find a larger R-band (which encompasses the H$\alpha$ line) transit depths in HAT-P-33b and CoRoT-12b and this result may indicate possible atmospheric escape. Follow-up photometry and high-resolution spectroscopy observations are encouraged to confirm all the observed transit depth variations since they are only seen at 2-4.6$\sigma$. Our calculated physical parameters agree with previous studies within 1$\sigma$ with a few exceptions (Tables \ref{tb:pars_1_light}--\ref{tb:pars_2_light}). For the exoplanets HAT-P-12b, HAT-P-37b, WASP-2b, WASP-24b, WASP-80b, and XO-3b we are able to refine their orbital periods from previous work (Tables \ref{tb:pars_1_light}--\ref{tb:pars_2_light}). 



\section*{Acknowledgments}

J. Turner, R. Leiter, and R. Johnson were partially supported by the NASA's Planetary Atmospheres program. J. Turner and R. Leiter were also supported by the The Double Hoo Research Grant. J. Turner was also partially funded by the National Science Foundation Graduate Research Fellowship under Grant No. DGE-1315231. 

We would like thank Robert T. Zellem for his help with observations. This research has made use of the Exoplanet Orbit Database \citep{Wright2011exo}, Exoplanet Data Explorer at exoplanets.org, Exoplanet Transit Database, Extrasolar Planet Transit Finder, NASA's Astrophysics Data System Bibliographic Services, and the International Variable Star Index (VSX) database, operated at AAVSO, Cambridge, Massachusetts, USA. This research has also made use of the NASA Exoplanet Archive, which is operated by the California Institute of Technology, under contract with the National Aeronautics and Space Administration under the Exoplanet Exploration Program. 
    


\bibliographystyle{mnras}
\bibliography{reference_new.bib}

\label{lastpage}
\end{document}